\documentclass[12pt]{article}
\usepackage[dvips]{graphicx}
\usepackage{amssymb}
\usepackage{amsmath}
\usepackage{epsfig}
\usepackage{cite}
\usepackage{hyperref}
\usepackage{bbold}
\usepackage{multirow}
\usepackage{verbatim}
\usepackage{enumitem}
\usepackage{color}
\usepackage{tabularx}

\numberwithin{equation}{section}
\numberwithin{table}{section}

\def\beq{\begin{equation}}
\def\eeq{\end{equation}}
\def\be{\begin{equation}}
\def\ee{\end{equation}}
\def\bea{\begin{eqnarray}}
\def\eea{\end{eqnarray}}

\DeclareRobustCommand{\SkipTocEntry}[4]{}
\RequirePackage{color}

\newcommand{\cV}{\mathcal{V}}

\newcommand{\bbZ}{\mathbb{Z}}
\newcommand{\bbR}{\mathbb{R}}

\newcommand{\tK}{{\widetilde{K3}}}

\begin{document}

\begin{titlepage}
\begin{center}
\rightline{\small }

\begin{flushright}
\end{flushright}

\vskip .1cm

{\Large \bf The Tadpole Problem }
\vskip 1.2cm

{ Iosif Bena$^{a}$, Johan Bl{\aa}b{\"a}ck$^{b}$, Mariana Gra\~na$^{a}$ and
Severin L\"ust$^{c,d}$ }
\vskip 0.1cm
{\small\it  $^{a}$ Institut de Physique Th\'eorique,
Universit\'e Paris Saclay, CEA, CNRS\\
Orme des Merisiers \\
91191 Gif-sur-Yvette Cedex, France} \\
\vskip 0.1cm
{\small\it  $^{b}$ Dipartimento di Fisica, Universit\`a di Roma ``Tor Vergata" \& INFN - Sezione di Roma2 \\
Via della Ricerca Scientifica 1, 00133 Roma, Italy} \\
\vskip 0.1cm
{\small\it  $^{c}$ Jefferson Physical Laboratory, Harvard University, Cambridge, MA 02138, USA }
\vskip 0.1cm
{\small\it  $^d$ Centre de Physique Th\'eorique, Ecole Polytechnique, CNRS \\
91128 Palaiseau Cedex, France} \\
\vskip 0.8cm

{\tt }

\end{center}

\vskip 1cm

\begin{center} {\bf Abstract }\\
\end{center}

We examine the mechanism of moduli stabilization by fluxes in the limit of a large number of moduli. We conjecture that one cannot stabilize all  complex-structure moduli in F-theory at a generic point in moduli space (away from singularities) by fluxes that satisfy the bound imposed by the tadpole cancelation condition. More precisely, while the tadpole bound in the limit of a large number of  complex-structure moduli goes like $1/4$ of the number of moduli, we conjecture that the amount of charge induced by fluxes stabilizing all moduli grows faster than this, and is therefore larger than the allowed amount.
Our conjecture is supported by two examples:  $K3 \times K3$ compactifications, where by using evolutionary algorithms
we find that moduli stabilization needs fluxes whose induced charge is 44\% of the number of moduli, and Type IIB compactifications on $\mathbb{CP}^3$, where the induced charge of the fluxes needed to stabilize the D7-brane moduli is also 44\% of the number of these moduli. Proving our conjecture would rule out de Sitter vacua obtained via antibrane uplift in long warped throats with a hierarchically small supersymmetry breaking scale, which require a large tadpole.

\vspace{0.2cm}

\noindent

\vfill

\today

\end{titlepage}


\tableofcontents

\newpage

\section{Introduction}

Despite the existence of a large number of ways to compactify String Theory to four dimensions, the requirement that the resulting four-dimensional universe have no massless scalars severely constrains the possibilities. These massless scalars come from the complex-structure and the K\"ahler moduli of the compactification manifold, as well as from D-brane moduli, and to fix them perturbatively one has to turn on nontrivial fluxes on the compactification manifold. However, the effect of these fluxes is not innocuous. They contribute non-trivially to various brane tadpoles, which have to remain zero in a compact manifold. Hence, these tadpole contributions have to be canceled by some other ingredients.

It is fair to say that as a result of all these constraints, constructing explicitly any type of flux compactification with stabilized moduli is a rather artisanal process. The purpose of the paper is to understand how the contribution of the fluxes to the global tadpole scales with the number of moduli that need to be stabilized, and to argue that a large number of moduli can never be stabilized within tadpole constraints.

The easiest language in which one can make these arguments is that of F-theory, where the only negative contribution to the D3 tadpole comes from the Euler characteristic, $\chi_{CY4}$, of the compactification four-fold. This Euler characteristic is related in turn to the number of moduli that has to be stabilized using fluxes. One can obtain a large $\chi_{CY4}$ because of a large $h^{3,1}$, or a large $h^{1,1}$, or both. However, the latter correspond to K\"ahler moduli, and one does not expect in general to be able to stabilize too many of these moduli.\footnote{A quick way to see this is to remember that K\"ahler moduli are stabilized by non-perturbative effects on D7 branes wrapping the corresponding 4-cycles; however these branes also contribute to the D7 tadpole, and their contribution needs to be canceled by O7 planes, whose number is finite.} Hence, the only large-Euler-number compactifications that can be in principle stabilized will have a large number of complex-structure moduli, and a negative contribution to the tadpole given by
\beq
-Q_{\rm D3}^{\rm neg} =  \frac{\chi_{CY4}}{24} \sim \frac{h^{3,1}}{4}\label{F-tad-intro}  ~\quad {\rm at ~large ~} h^{3,1} \ .
\eeq
The $h^{3,1}$ complex-structure moduli are stabilized by four-form fluxes  \cite{Dasgupta:1999ss}, which give a positive contribution to the tadpole.

There is a clear link between the F-theory language and the standard language of Type IIB compactifications, and we will translate our arguments in that language as well.

We make three conjectures about the growth of the flux contribution to the D3 tadpole with the number of moduli that have to be stabilized by fluxes.

\subsection{Three conjectures}

\medskip \hypertarget{conjecture1} 
\noindent  {\bf 1. The Tadpole Conjecture:} \label{conjecture1} \medskip \\
{\it The fluxes which stabilize a large number, $h^{3,1}$, of complex-structure moduli of a Calabi-Yau four-fold in a smooth F-theory compactification will have a positive contribution to the D3 brane tadpole that grows at least linearly with $ h^{3,1}$}:
\beq
Q_{\rm D3}^{\rm stabilization} > \alpha \times h^{3,1}   ~\quad {\rm at ~large ~} h^{3,1} \,
\label{conj1-intro}
\eeq
{\it where $\alpha$ is a constant.}

We will call $\alpha$ the flux-tadpole constant. We will present analytic arguments as well as evidence based on differential evolution algorithms for this Tadpole Conjecture, and point out some ways in which one could try to prove it. Note that this conjecture implies that one cannot stabilize a growing number of moduli with a finite number of fluxes, whose tadpole would stay $O(1)$.

Based on our existing evidence, it also appears that the flux-tadpole proportionality constant, $\alpha$ is bounded below. This leads us to

\medskip \hypertarget{conjecture2} 
\noindent  {\bf 2. The Refined Tadpole Conjecture:}  \medskip \\
{\it The flux-tadpole constant, $\alpha$, is bounded below:}
\beq \label{alpha}
\alpha> \frac13  \,.
\eeq

In Section \ref{K3xK3} we will explain how we arrive at this lower bound, and in Section \ref{sec:D7} we present further supporting evidence by using a compactification studied in \cite{Collinucci:2008pf}. The very recent example of \cite{Braun:2020jrx} on the sextic CY  four-fold gives additional support. We hope to present further evidence for Conjecture 2 in upcoming work.

When translating our conjectures in the language of Type IIB compactifications, the F-theory complex-structure moduli become a combination of D7 brane moduli and Type IIB complex-structure moduli. Stabilizing these moduli also comes with a price. The complex-structure moduli are stabilized by  three-form fluxes \cite{Grana:2001xn, Giddings:2001yu,Gukov:1999ya}, and these fluxes contribute nontrivially to the D3 tadpole. The D7 moduli are stabilized by introducing worldvolume two-form fluxes, and these fluxes also give rise to finite contributions to the D3 tadpole.

From a Type-IIB perspective the D3 tadpole budget needed for stabilizing the D7 moduli and the D3 tadpole budget needed for stabilizing Type IIB complex-structure moduli appear to come from very different physics, which leads us to formulate two conjectures:
\newpage
\medskip \hypertarget{conjecture3a} 
\noindent {\bf 3a. The Type IIB Tadpole Conjecture A:} \medskip \\
{\it The fluxes which stabilize the $h^{2,1}$ complex-structure moduli of a smooth Calabi-Yau three-fold in a Type IIB  compactification will give a positive contribution to the D3 brane tadpole that for large $ h^{2,1}$ grows at least linearly with the number of stabilized moduli: }
\beq
Q_{\rm D3}^{(2,1)\,\rm  stabilization} > \alpha \times h^{2,1}\,. \label{conj3a}
\eeq

\medskip \hypertarget{conjecture3b} 
\noindent {\bf 3b. The Type IIB Tadpole Conjecture B :\label{conjecture3b}} \medskip \\
{\it The worldvolume fluxes which stabilize the $n_7$ moduli of a collection of D7 branes wrapping a four-cycle have a positive contribution to the D3 tadpole that for large $n_7$ grows at least linearly with the number of D7 moduli}
\beq
Q_{\rm D3}^{\rm D7\,stabilization} > \alpha \times n_7 \,. \label{conj3b}
\eeq

These two conjectures are not independent. In fact, Conjecture~\hyperlink{conjecture3a}{3a} together with Conjecture~\hyperlink{conjecture1}{1} implies Conjecture~\hyperlink{conjecture3b}{3b}, and viceversa. Hence, they can be thought of as two equivalent facets of the same conjecture.

Conjecture 3a is based on the similarity between the Type IIB and the F-theory equations governing the stabilization of moduli and the contribution of the fluxes to the D3 brane tadpole, and also of the similarity between flux compactifications and bubbling solutions that will be described below.

It is important to note the essence of these conjectures: the fact that the D3 tadpole sourced by the fluxes needed to stabilize the moduli grows {\em linearly} with the number of moduli is not directly linked with the ultimate size of the cycles that end up stabilized. This growth is rather a feature of the topological structure of the compactification manifold and of the fact that the D3 charge dissolved in fluxes cannot be negative in any region of the compactification manifold.  In particular, if a certain flux configuration stabilizes all the moduli, increasing the value of all the fluxes by a factor  $k$  increases the overall D3 tadpole by a factor of $k^2$; however, the dependence of this tadpole on the number of moduli to be stabilized still remains linear.\footnote{We thank A. Braun and J. Moritz for interesting discussions on this point.}

\subsection{Three consequences}

There are three important consequences of the physics behind  our conjectures:

$\bullet $ The first is ruling out all stable F-theory compactifications {on smooth manifolds}  with a large number of moduli. This is because  the D3 charge induced by the fluxes needed to stabilize the moduli \eqref{conj1-intro} is larger than the tadpole bound \eqref{F-tad-intro}.
There exist Calabi-Yau four-folds with a very large number of cycles and a very large Euler characteristic. The number of possible F-theory flux compactifications corresponds to the various ways of putting fluxes on the cycles of such four-folds, and has been estimated to be of order $10^{272000}$ \cite{Taylor:2015xtz}. Our conjectures  indicate that none of these vacua has stabilized moduli.

 $\bullet $ The second is ruling out AdS compactifications with scale separation that combine fluxes and non-perturbative ingredients \`a la KKLT  \cite{Kachru:2003aw}.
 These scale-separated AdS vacua can only be built when the vacuum expectation value of the flux superpotential is very small, comparable to the non-perturbative contribution.
Such small values of the superpotential can only be achieved with a large number of moduli \cite{Denef:2004ze}.\footnote{By playing with instanton contributions to the prepotential,  one can actually engineer small values of the superpotential at smaller values of the tadpole than the statistics arguments suggest \cite{Demirtas:2019sip}, though still relatively large tadpoles are needed \cite{Demirtas:2020ffz, Blumenhagen:2020ire}.} Our conjectures imply that these compactifications have leftover unfixed complex-structure moduli even prior to considering non-perturbative contributions that stabilize K\"ahler moduli. This result is consistent with the swampland-based {\em AdS Distance Conjecture} against the existence of vacua with scale separation between the AdS scale and the scale of the compactification manifold \cite{Lust:2019zwm} and also with the recent bootstrap-based arguments against such compactifications \cite{Alday:2019qrf}.

$\bullet$ The third is ruling out de Sitter vacua obtained by a small uplift of an AdS vacuum  \cite{Kachru:2003aw}. This uplift is realized by adding antibranes at the bottom of a long warped Klebanov-Strassler-like throat \cite{Klebanov:2000hb}, and in \cite{Bena:2018fqc} we have found that even a single antibrane destabilizes this throat unless one of the fluxes exceeds a certain lower bound. As we will discuss in detail in Section \ref{KS}, when combining this with requiring also a large hierarchy between the bottom of the throat and the rest of the compactification manifold (necessary to avoid the destabilization of the other moduli) one finds that the contribution of the fluxes in the Klebanov-Strassler throat to the D3 brane tadpole should be at least of order 500.
Our conjectures rule out the possibility of canceling such a large tadpole while stabilizing all the other moduli of the compactification, and hence it rules out de Sitter vacua obtained via uplifting using antibranes in long warped throats.

\subsection{Three supporting arguments}

 $\bullet$  The first two conjectures indicate that there do not exist any smooth F-theory compactification \footnote{Singularities give rise to non-Abelian gauge groups which carry additional massless fields.} with a large Euler characteristic and stabilized moduli. It is interesting to understand above what values of the Euler characteristic these conjectures  start applying. There are certainly F-theory compactifications with a very small number of complex-structure moduli where all the moduli can be stabilized \cite{Denef:2005mm,Honma:2017uzn}. However, as we will see in detail in Section \ref{K3xK3}, even in a smooth $K3 \times K3$ F-theory compactification, the minimal amount of fluxes needed to stabilize the moduli already gives a positive induced charge that is larger than what can be absorbed from the Euler characteristic of this space. In particular, this contribution is almost as large as the number of stabilized moduli divided by two.

Indeed, if one considers random configurations of fluxes that stabilize all the complex-structure moduli, most of these configurations have a tadpole that is of order  $h^{3,1} \times h^{3,1}$. This makes sense, since the intersection matrix is an   $h^{3,1}$ by   $h^{3,1}$ matrix with numbers of order one in every entry, and the fluxes are also numbers of order one.

However, using an intelligent computer search based on evolutionary algorithms,%
\footnote{Evolutionary or genetic algorithms were also used to explore the string landscape and flux compactifications in \cite{Blaback:2013ht, Damian:2013dq, Damian:2013dwa, Blaback:2013fca, Blaback:2013qza, Abel:2014xta, Ruehle:2017mzq, Cole:2019enn,AbdusSalam:2020ywo,CaboBizet:2020cse}.
Another computer aided search of flux vacua constrained by tadpole cancellation was performed in \cite{Betzler:2019kon}.
For a general overview of data science and machine learning techniques in the context of String Theory see for example \cite{Ruehle:2020jrk} and references therein.
}
 one can try to build configurations with smaller tadpole. Our search has actually produced some flux configurations whose induced D3 charge is close to the tadpole bound (though always larger!), and they all satisfy conjectures~\hyperlink{conjecture1}{1} and \hyperlink{conjecture2}{2}. These configurations are very rare, and take longer and longer to find as one decreases the tadpole. However, we could never lower the tadpole of the fluxes that stabilized all moduli below 25, which corresponds to a flux-tadpole constant, $\alpha$, of order $44\%$, largely above the bound of Conjecture~\hyperlink{conjecture2}{2}. We will provide several other examples with less moduli which confirm this linear growth in an upcoming paper \cite{algorithm}.

Our example illustrates that even for spaces with a relatively small Euler characteristic one cannot stabilize all the moduli that can be stabilized by fluxes, while maintaining a small induced D3-charge.

 $\bullet$ One can also bring evidence in favor of Conjecture~\hyperlink{conjecture3b}{3b} by investigating a particular example of D7 moduli fixing in a Type IIB compactification on a six-dimensional manifold $X$ \cite{Collinucci:2008pf}.
The Euler characteristic of a smooth four-cycle wrapped by the D7 branes is given by the adjunction formula \eqref{adj}, 
and one can see that the only way to obtain a very large negative D3 tadpole contribution from the D7 branes is to wrap them on a curve with a very large self-intersection number.
However, this curve always comes with a large number of D7 moduli, which in the large Euler number limit grows linearly with the Euler number. In Section \ref{sec:D7} we will discuss in detail the example  worked out in \cite{Collinucci:2008pf} and find that $\alpha=1$, again largely above the bound of Conjecture~\hyperlink{conjecture2}{2}.

 $\bullet$ There is another physically interesting system where the contribution to the tadpole of the fluxes needed to stabilize the size of various topologically-nontrivial cycles has important physical consequences: The construction of bubbling horizonless microstate geometries that have the same mass, charge and angular momenta as a black hole. To build these solutions one adds fluxes on the topologically-nontrivial 2-cycles of a multi-center Gibbons-Hawking space \cite{Bena:2005va,Berglund:2005vb}, and these fluxes prevent the cycles from collapsing. If one wants to construct a horizonless geometry corresponding to a black hole of charge $Q$, the total contribution to the tadpole of the fluxes in the geometry has to be equal to $Q$, and the multiple possible ways of putting fluxes on various cycles correspond to different microstates of the corresponding black hole.

If one proceeds na\"ively and one does not impose the absence of closed timelike curves, one can easily obtain a very large number of geometries for a given total black hole charge. These geometries easily over-count the entropy of the black hole. However, if one asks for no closed timelike curves, one finds that most of these na\"ive geometries have to be discarded, and that very few regular geometries remain. The geometries with closed timelike curves have in general large positive tadpole contributions in some regions, and large negative contributions in others, such that the total tadpole stays finite. In contrast, the regular geometries have a local tadpole contribution that is positive throughout the solution.

One can then estimate the growth of the tadpole sourced by the fluxes with the number $n$ of stabilized cycles, to be proportional to $n^2$ \cite{Bena:2006is}. As we will see in Section \ref{K3xK3}, this is the same growth that we find for the most generic distribution\footnote{It is only by choosing very special fluxes that one can get this growth to be linear. It may be possible that with enough searching one could find some bubbling solutions where the charges sourced by the fluxes also grow linearly with the number of stabilized cycles.} of fluxes on $K3 \times K3$.

However, it is clear that one cannot stabilize a growing number of cycles using fluxes whose tadpole contribution stays of order one. If this were possible, one could construct horizonless bubbling solutions with an arbitrarily large number of cycles and the same charge, which would over-count the black hole entropy and directly disprove the AdS-CFT correspondence. This fact lends further support to our conjectures. 

Besides the two examples we discuss in detail in this paper, there are several other examples in the literature where a relatively large number of moduli have been stabilized with fluxes and the tadpole contribution of these fluxes has been computed \cite{Collinucci:2008pf,Giryavets:2003vd, Denef:2004dm, Demirtas:2019sip,Braun:2020jrx}. All these examples are consistent with the refined tadpole conjecture, as one can see from Table~\ref{tab:examples}.

\begin{table}[htb]
\centering
\setlength\extrarowheight{1pt} 
\begin{tabularx}{\textwidth}{|X|c|c|c|c|} \hline 
&&&& \\[-0.64cm]
Description  \vspace*{8pt} & $n_\mathrm{mod.}$ & $Q_{\rm D3}^\mathrm{flux}$ & $\alpha = {\displaystyle \frac{Q_{\rm D3}^{\mathrm{flux}}}{ 
n_\mathrm{mod.}}}$ & Ref. \\
\hline \hline
Stabilization of D7-brane moduli in the IIB limit of F-theory with a $\mathbb{CP}^3$
 base (see Section~\ref{sec:D7}) & $n_7 = 3728$ & $1638$ & $0.44$ &  \hbox{\cite{Collinucci:2008pf}} \\ \hline
\multirow{2}{\hsize}{A Type-IIB compactification at a highly symmetric point in moduli space (see also \cite{Denef:2004dm})} & $h^{2,1} = 128$  & $48$ & $0.38$ & \hbox{\cite{Giryavets:2003vd}} \\ 
 & $h^{2,1} = 272$ & $124$ & $0.46$ & \hbox{\cite{Demirtas:2019sip}} \\[5pt]  \hline
F-theory on the sextic Calabi-Yau four-fold with fluxes on algebraic cycles  & $h^{3,1} = 426$ & $775/4$ & $0.45$ & \hbox{\cite{Braun:2020jrx}} \\ \hline
M-theory on smooth $K3 \times K3$ using an evolutionary algorithm
(Section~\ref{K3xK3}) & $57$ & $25$  & $0.44$ &  \\ \hline
\end{tabularx}
\caption{Examples for flux compactifications at a large number of moduli.
We give the number of moduli which are stabilized by fluxes, $n_\mathrm{mod.}$, the charge induced by the fluxes which stabilize these moduli, $Q_{\rm D3}^\mathrm{flux}$,and the corresponding value of $\alpha$ in the tadpole conjectures.}
\label{tab:examples}
\end{table}

In Section \ref{sec:fluxcompact} we discuss the basic features of F-theory and Type IIB flux compactifications, concentrating in Section \ref{sec:modstab} on the mechanism of moduli stabilization. In Section \ref{K3xK3} we discuss the example of $K3 \times K3$ and show the results of the evolutionary algorithms. We only give a qualitative account of the algorithms, relegating the technical details to a subsequent paper \cite{algorithm}. In Section \ref{sec:D7} we discuss the Type IIB example worked out in \cite{Collinucci:2008pf}.  In Section \ref{KS} we show that the bound on $M$ gives a gives a lower bound of 500 on the tadpole of a KS throat with antibranes, and we argue that one cannot cancel such a large tadpole in a compactification with stabilized moduli.  We present some conclusions and future directions in Section \ref{sec:disc}.

\section{Flux compactifications}
\label{sec:fluxcompact}

Fluxes in warped compactifications have to satisfy  tadpole cancelation conditions arising from Bianchi identities.
These identities lead to local equations, which enforce a profile for the warp factor in terms of the fluxes and local sources. Here we are interested in the global tadpole conditions that arise from integrating these local equations over the whole internal manifold. In these global equations the warp factor disappears, and one obtains conditions enforcing the cancelation of the total charge sourced by branes and orientifolds and induced by the fluxes and the curvature.

In Type IIB compactifications with 3-form fluxes, the cancellation condition of the D3-brane tadpole is
\beq \label{tadpoleIIB}
{\color{black} \frac12} \int_X F_3 \wedge H_3- {\color{black} \frac{1}{2}} \int_S \bigl[\rm{Tr}(F \wedge F) - \rm{Tr} F \wedge \rm{Tr} F\bigr]
 + Q_3^{\rm{loc}} =0 \,.
\eeq
 where $X$ is the double cover of the six-dimensional manifold and $S$ denotes the four-cycle wrapped by D7-branes with worldvolume flux $F$. In these compactifications the fluxes that solve the equations of motion are self-dual\footnote{In this convention, bulk three-form fluxes satisfy $H_3=e^{\phi} *_6 F_3$ \cite{Dasgupta:1999ss,Grana:2001xn,Giddings:2001yu} and D7 worldvolume fluxes satisfy $F_2 = -*_S F_2$.} and their contribution to the tadpole is always positive. This positive charge has to be canceled by localized negative-charge objects, of which there are three kinds:\\
$\bullet$ O3 planes\\
$\bullet$ D7 branes wrapping 4-cycles of Euler number $\chi (D7_i)$\\
$\bullet$ O7 planes wrapping 4-cycles of Euler number $\chi (O7_i)$\\
The D3-charge of the D3 brane sources and of these localized negative-charge sources is\footnote{We are using conventions for type IIB compactifications with orbifolds in which a mobile D3-brane is counted together with its image brane {\color{black} as one brane} and corresponds to $N_{D3}={\color{black} 1}$.
{There exists another convention in the literature where each pair of a brane with its image counts as $N_{D3} = 2$. This convention comes at the price of an additional factor of $2$ when translating the F-theory charge $\chi(CY_4)/24$ into the IIB charge $Q_3^\mathrm{loc}$, and we will not use it in this paper.} \label{foot:chargeIIB}}  
\beq\label{IIBcharge}
 Q_3^{\rm{loc}}= -{\color{black} \frac14} N_{O3} - \frac{1}{24} \Bigl[\chi(D7) + {\color{black} 2} \chi(O7)\Bigr]  + N_{D3}\,.
\eeq

Here we are interested in compactifications with large negative contributions to the tadpole, which are the ones susceptible to give scale-separated AdS and de Sitter vacua.  Since one cannot obtain a large negative charge from O3-planes, the compactification should have D7-branes as well, and as such it is more convenient to use the language of F-theory, where D7-branes and O7-plane sources are nicely encoded in the geometry of a Calabi-Yau four-fold, and their negative contribution to the D3-brane tadpole is captured by the Euler characteristic of the  whole manifold.  The NSNS and RR 3-form field fluxes, as well as the D7 fluxes are combined into a 4-form field-strength $G_4$, and the D3-brane tadpole cancelation condition is simply\footnote{There can be additional O3-planes which are not geometrized, but their contribution to the tadpole parametrically smaller than $\chi/24$ in the limit that we are interested in, and thus we do not include them here.}
\beq
\label{Ftadpole}
\frac12 \int_{CY_4} G_4 \wedge G_4+N_{D3}-\frac{\chi(CY_4)}{24}=0 \,.
\eeq
In the description of F-theory via M-theory, this is expression  is the tadpole-cancelation condition for M2-brane charge. The Euler number of the 4-fold is related to its Hodge numbers by
\beq \label{chiCY4}
\chi(CY_4)=6(8+h^{1,1}+h^{3,1}-h^{2,1}) \,,
\eeq
where we have used the fact that
\beq
h^{2,2}= 2(22+ 2 h^{1,1}+2h^{3,1}-h^{2,1}) \,.
\eeq
The Hodge numbers of the four-fold are in turn related to the dimension of the moduli space of F-theory compactifications. There are  $h^{1,1}$ K\"ahler moduli, and $h^{3,1}$ complex-structure moduli. We list all the moduli in Appendix \ref{App:moduli}, where we also give their Type IIB origin.

If one wants to construct Calabi-Yau 4-fold compactifications with stabilized moduli and a large Euler number\footnote{The Euler number of Calabi-Yau 4-folds can be at large as 1\,820\,448 \cite{Klemm:1996ts}.}, it is not hard to see  that most of the right-hand side of equation \eqref{chiCY4} has to come from
$h^{3,1}$. This is because increasing $h^{2,1}$ decreases the Euler number, and the $h^{1,1}$ are K\"ahler moduli that are much harder to stabilize (there is no known perturbative mechanism for this). Hence, in the large-Euler-number regime, \eqref{chiCY4} implies
\beq \label{chilargeh31}
\frac{\chi(CY_4)}{24}\sim \frac{h^{3,1}}{4} \ .
\eeq

In the Type IIB limit the total negative charge, \eqref{IIBcharge}, cannot be expressed only in terms of the topological data of the 3-fold.
Nevertheless, one can use the Lefshetz fixed-point theorem to constrain at least the contribution of the orientifold action to the D3 tadpole \eqref{IIBcharge},\footnote{We thank C.~Vafa for pointing this out to us.}
\begin{equation}
N_{O3} + \chi(O7) = 2 \left[2 + (h^{1,1}_+ - h^{1,1}_-) - (h^{2,1}_+ - h^{2,1}_-)\right] < 4 + 2 (h^{1,1} + h^{2,1}) < 1008 \,.
\label{O7bound}
\end{equation}
where the last inequality is based on the Calabi-Yau 3-folds in the Kreuzer-Skarke list \cite{Kreuzer:2000xy}. Hence the absolute value of the maximal negative contribution to the D3 tadpole that can come from O7 planes is 
\begin{equation}
\left| Q^{-}_{\displaystyle \chi_{\rm O7}} \right| = {\chi(O7) \over {\color{black} 12}} <{\color{black} 84} \,.
\label{O7bound2}
\end{equation}
{\color{black}In the particular situation when all D7-branes are on top of the orientifold planes, we have
\begin{equation}
\left|Q^{\rm{loc}}_{\chi({\rm D7 \, on \, top \, of \, O7})}\right|= \frac14 N_{O3} + \frac{1}{24}  \Bigl[4 \chi(O7) + 2 \chi(O7)\Bigr]   =\frac14 (N_{O3} + \chi(O7) ) < 252 \ .
\end{equation}
However, for more generic D7 brane shapes,}  the negative D3 tadpole contribution from D7 branes does not have such bounds.
Neverthless, in the large $\chi(D7)$ limit we can use the fact that the number of D7 moduli and $\chi(D7)$ are both proportional to the triple intersection number of the four-cycle on which the D7 branes are wrapped.
Calling this four-cycle $S$ we thus have
\beq\label{eq:D7moduli}
\left| Q^{-}_{\displaystyle \chi_{\rm D7}} \right| = \frac{\chi(S)}{24} \sim \frac{1}{24} \int_X S^3 \sim \frac{ n_\text{D7 moduli}}{4} \,
\eeq
where the relation between $\chi(S) $ and $\int S^3$ is the large-$\chi(S) $ limit of the adjunction formula
\beq \label{adj}
\chi(S)=\int_X \left(S^3 + c_2(X) S\right)
\eeq
and the relation between the number of D7 moduli and the triple-intersection number is probably best known from the MSW microscopic calculation of black-hole entropy \cite{Maldacena:1997de}. This relation is also consistent with the F-theory large-Euler-characteristic relation \eqref{chilargeh31}, and this illustrates the fact that from a Type-IIB perspective the only negative contribution to the D3 tadpole that can grow arbitrarily comes from D7 branes wrapping four-cycles with arbitrarily-large self-intersection numbers.

\section{Moduli Stabilization and Tadpole Cancelation}
\label{sec:modstab}

The main reason why fluxes are a necessary ingredient in compactifications is their ability to stabilize moduli: Fluxes create a potential for the moduli because the ten-dimensional supergravity action has terms of the form $F \wedge * F$, and the Hodge star depends on the moduli of the compactification manifold. For a three-fold, the Hodge star acting on even forms depends on the K\"ahler moduli, while for odd forms it depends on the complex-structure moduli. Since in Type IIB String Theory all fluxes are odd forms, only complex-structure moduli can be stabilized by fluxes. There are  $h^{2,1}(CY_3)$ such moduli, which combine to the axion-dilaton and the D7-brane moduli to give the $h^{3,1}(CY_4)$ complex-structure moduli of an F-theory compactification. In the language of F-theory these moduli are stabilized by four-form fluxes.

We very briefly recall  the mechanism of moduli stabilization in F-theory \cite{Dasgupta:1999ss,Giddings:2001yu}. Taking a basis for the integral cohomology four-forms
$\omega_A \in H^4(CY_4, \bbZ)$, $A={1,...,b^4}$, where $b^4=2+2 h^{3,1}+h^{2,2}$, the flux is
\beq
G_4=N^A \omega_A
\eeq
with $N^A \in {\mathbb Z}$. The flux-induced tadpole is
\beq
\frac12 \int_Z G_4 \wedge G_4= \frac12 N^A\,   \eta_{AB} \, N^B \,,
\eeq
where $\eta_{AB}$ is the intersection matrix. Since this matrix has indefinite signature, the contribution to the tadpole from a generic set of fluxes can a priori have any sign.
However, not any set of fluxes is allowed:
the flux-induced potential (ignoring numerical pre-factors) is \cite{Denef:2008wq},
\begin{equation}\label{eq:potential}
V = \frac{1}{\cV^3} \int_{Z} G_4 \wedge \star G_4 - G_4 \wedge G_4 \,,
\end{equation}
where $\cV = \tfrac{1}{4!}\int_{CY_4} J^4$ denotes the volume of $CY_4$. In this expression, the second term is an on-shell contribution, obtained using the Bianchi identity for $G_4$.
This potential is positive semi-definite and has Minkowski minima at values of the moduli where $G_4$ is self-dual,
\beq \label{sd4}
*G_4 =G_4  \ .
\eeq
To analyze the implications of this condition, we decompose the middle cohomology into a selfdual and anti-selfdual part,
\begin{equation}
H^4 = H^4_+ \oplus H^4_- \,.
\end{equation}
Remember that in a four-fold, a $(p,q)$-form $\alpha^{(p,q)}$ is  primitive if
\begin{equation}
J^{5 - p - q} \wedge \alpha^{(p,q)} = 0 \,.
\end{equation}
Primitive four-forms satisfy
\begin{equation}
\star \alpha^{(p,4-p)} = (-1)^p \alpha^{(p,4-p)} \,,
\end{equation}
therefore we have
\begin{equation}\begin{aligned}
H^4_+ &= H^{4,0} \oplus H^{2,2}_{(0)} \oplus H^{2,2}_{(2)} \oplus H^{0,4} \,, \\
H^4_- &= H^{3,1} \oplus H^{2,2}_{(1)} \oplus H^{1,3} \,,
\end{aligned}\end{equation}
where $H^{p,p}_{(l)}$ denotes the Lefshetz decomposition of $H^{p,p}$ :
\begin{equation} \label{Lefshetz}
H^{p,p}_{(l)} = J^l \wedge H^{p-l, p-l}_{(0)} \,,
\end{equation}
and $H^{p-l, p-l}_{(0)}$ is spanned by primitive forms.
With this knowledge the potential \eqref{eq:potential} can be rewritten as
\begin{equation} \label{pot}
V = - \frac{1}{\cV^3} \int_{CY_4} 4 G^{3,1} \wedge G^{1,3} + 2 J \wedge G^{1,1}_{(0)} \wedge J \wedge G^{1,1}_{(0)} \, .
\end{equation}
As explained in \cite{Haack:2001jz}, this potential can be obtained using the two superpotentials\footnote{Strictly speaking $\hat W$ is not a superpotential, as it is real. It can nevertheless  be used to derive the primitivity flux condition, which in four dimensions should be interpreted as a D-term, rather than an F-term constraint.}
\begin{equation}
W = \int_{CY_4} \Omega \wedge G_4 \,,\qquad \hat W = \frac14 \int_{CY_4} J \wedge J \wedge G_4
\end{equation}
and the K\"ahler potential
\begin{equation}
K = - \log\left[\int_{CY_4} \Omega \wedge \bar \Omega \right] - 3 \log\left[\frac{1}{4!}\int_{CY_4} J^4\right] \,.
\end{equation}
There is a supersymmetric Minkowski minimum iff
\begin{equation} \label{susyconditions}
D_\alpha W = 0 \,,\quad W=0 \,,\quad  D_A \hat W =  0 \,,\quad \hat W =  0 \,,
\end{equation}
where $D_\alpha W = \partial_\alpha + (\partial_\alpha K) W$ denotes the K\"ahler-covariant derivative with respect to the complex-structure moduli, $z^\alpha$ ($\alpha = 1, \dots, h^{3,1}$) and $D_A$ ($A = 1, \dots, h^{1,1}$) denotes the covariant derivative with respect to the K\"ahler moduli.
The first condition implies that $G^{3,1} = 0$, the second implies that $G^{4,0} =0$, the third implies that $ G^{2,2}_{(1)} = 0$ and the last one implies that $G^{2,2}_{(2)}=0$.
For manifolds of strict SU(4) holonomy (as opposed to $K3 \times K3$, for example) the primitivity condition of the four-form flux is automatic for fluxes that have one leg on the elliptic fiber,
such as the F-theory uplift of $H_3$ and $F_3$. The last condition in \eqref{susyconditions} becomes thus trivial and does not help in stabilizing K\"ahler moduli.

Since the potential \eqref{pot} is positive semi-definite, it is easy to see that there is a Minkowski minimum even if $G^{0,4}$ or $G^{2,2}_{(2)}$ are non-zero.
Such minima are non supersymmetric. In summary, $G_4$ must be a primitive (2,2)-form in supersymmetric solutions, and is allowed to have a (0,4)
component and/or a non-primitive $G^{2,2}_{(2)}$ piece in non-supersymmetric Minkowski vacua.
%
%

The condition $G^{3,1}=0$ gives a priori $h^{3,1}$ independent equations for the complex-structure moduli which are generically fixed.
However, the number of independent conditions depends on which fluxes, $N^A$, are turned on. Intuitively, one might say that one needs to turn on a flux for each cycle corresponding to a modulus that needs to be stabilized. However, this intuition might be a bit na\"ive. Also, the question as such is not well posed since, depending on the geometry of the moduli space, a single unit of flux can give rise to several equations.
However, as we have shown, the flux contribution to the tadpole is always positive, no matter how simple or complicated the $N^A$ are. Furthermore, if the $N^A$ are too generic, it is also possible that the potential has no minimum. A well posed, highly non trivial question is what is the minimum charge in the fluxes required to stabilize all moduli at a non-zero value, such that  shrinking-cycle singularities are avoided.  Based on the intuition from black-holes explained in the Introduction, and supported by the example in the next section, we conjecture that for a large number of moduli this charge satisfies
\beq
\frac12 \int_{CY_4} G_4 \wedge G_4 \Bigl.\Bigr|_{\rm all \ moduli \ stabilized} \equiv Q_{\rm D3}^{\rm stabilization} > \alpha h^{3,1} \label{conj1}
\eeq
with the flux-tadpole constant  $\alpha$ larger than $1/3$.

On the other hand, in order to satisfy the tadpole cancelation condition \eqref{Ftadpole} in the limit of a large number of complex-structure moduli, eq.~\eqref{chilargeh31} indicates that the flux-tadpole constant, $\alpha$, cannot be greater than  $1/4$. Hence this conjecture rules out F-theory compactifications with stabilized moduli when the Euler number is large.
We show supporting evidence for this conjecture in the next sections.

In F-theory the four-fold $CY_4$  is an elliptic fibration $T^2 \hookrightarrow CY_4 \rightarrow X$ over a complex three-dimensional base manifold $B$.
This four-fold describes a Type IIB compactification on $X$ with varying axio-dilaton $\tau$, given by the complex-structure modulus of the fiber.
The singularities of the fibration correspond to 7-branes, inducing monodromies of $\tau$.
In the orientifold or weak-coupling limit \cite{Sen:1996vd, Sen:1997gv} one recovers a standard Type IIB orientifold compactification on a Calabi-Yau three-fold $X$, which is the double cover of $X$, with D7-branes wrapping a 4-cycle $S$ of $X$.
In Appendix \ref{App:moduli} we summarize how the moduli of the F-theory four-fold, $Z$, map into those of the Type IIB compactification.
Here, we are mainly interested in the stabilization of complex-structure moduli.
In particular, the $h^{3,1}(CY_4)$ complex-structure moduli of $Z$ correspond not only to the IIB axio-dilaton, $\tau$, and to the $h^{2,1}_-(X)$ complex-structure moduli of $X$, but also to the $h^{2,0}_-(S)$ complex deformation parameters of the D7 branes:
\begin{equation}
h^{3,1}(CY_4) = h^{2,1}_-(X) + h^{2,0}_-(S) + 1 \,.
\end{equation}
The F-theory conjecture, \eqref{conj1}, thus leads to two conjectures: In the large $h^{2,1}_-(X)$ limit:
\beq \color{black}
{\color{black} \frac12} \int_X F_3 \wedge H_3|_{\rm all \, (2,1) \, moduli \, stabilized} \equiv Q_{\rm D3}^{\rm (2,1) \, stabilization} > \alpha h^{2,1}_- \,,
\eeq
and in the large $h^{2,0}_-(S)$ limit:
\beq 
- {\color{black} \frac{1}{2}}\int_S \bigl[\rm{Tr}(F \wedge F) - \rm{Tr} F \wedge \rm{Tr} F\bigr]|_{\rm all \, D7\, moduli \, stabilized} \equiv Q_{\rm D3}^{\rm D7 \, stabilization} > \alpha h^{2,0}_-(S) \,. \label{D7-conj}
\eeq
In practice, however, if $h^{3,1}(CY_4)$ is large, this will typically result in a large number of D7 moduli since the number of complex-structure moduli is bounded for the known Calabi-Yau three-folds. Thus, the Type IIB limit of \eqref{conj1} corresponds to \eqref{D7-conj}. We will discuss this further in Section \ref{sec:D7}, where we review the compactification analyzed in \cite{Collinucci:2008pf}, in which the D7 moduli cannot be stabilized within the tadpole limit.

\section{Moduli stabilization in $K3 \times K3$}
\label{K3xK3}

The first supporting evidence for our conjecture comes from F-theory compactifications on $CY_4 = K3 \times K3$ \cite{Dasgupta:1999ss,Aspinwall:2005ad,Braun:2008pz,Braun:2010ff}. 
This is a particularly nice example with a relatively large number of moduli and where one has all the relevant topological data. Here we analyze this compactification using the approach of \cite{Braun:2008pz,Braun:2010ff}.

\begin{subsection}{The goals}

Let us start by recalling a few relevant facts about $K3$.
There is a natural inner product of signature (3,19) on $H^2(K3, \bbZ)$,
\begin{equation} \label{K3prod}
(\alpha, \beta) = \int_{K3} \alpha \wedge \beta \,.
\end{equation}
We can introduce a basis $\alpha_i \in H^2(K3, \bbZ)$, $i = 1, \dots, 22$ such that the corresponding intersection matrix
\begin{equation}
d_{ij} = \int_{K3} \alpha_i \wedge \alpha_j
\end{equation}
is given by
\begin{eqnarray}
d_{ij} = U \oplus U \oplus U \oplus (-E_8) \oplus (-E_8) \, \quad {\rm with} \ \  U=\begin{pmatrix} 0 & 1 \\ 1 & 0 \end{pmatrix}
\end{eqnarray}
and $-E_8$ is minus the Cartan matrix of $E_8$.

The moduli space of $K3$ is the coset $\tfrac{O(3,19)}{O(3)\times O(19)}$, which has dimension $3\times 19=57$. A point in this moduli space corresponds to a choice of three 2-forms
\begin{equation}
\omega_{\hat \imath} \in H^2(K3, \bbR) \,,\qquad \hat \imath = 1,2,3 \,,
\end{equation}
of positive norm,~$(\omega_{\hat \imath}, \omega_{\hat \jmath}) = \delta_{\hat \imath \hat \jmath}$, which define a HyperK\"ahler structure on $K3$.
One can furthermore choose a real K\"ahler form $j = \sqrt{2 \mathrm{vol}(K3)} \, \omega_3$ and a holomorphic (2,0)-form $\Omega = \omega_1 + i \omega_2$.
The choice of $\omega_{\hat \imath}$ corresponds to the decomposition of $H^2(K3)$ into selfdual and anti-selfdual forms,
\begin{equation}
H^2(K3) =  H^2_+(K3) \oplus H^2_-(K3) \,,
\end{equation}
where $H^2_+(K3)$ is spanned by $\omega_{\hat \imath}$ and $H^2_-(K3)$ is its orthogonal complement.
We furthermore have
\begin{equation}
H^2_+ = H^{(2,0)} \oplus H^{(1,1)}_{(1)} \oplus H^{(0,2)}  \,,\qquad H^2_- = H^{(1,1)}_{(0)} \,,
\end{equation}
where the split of $H^{(1,1)}$ corresponds to a Lefshetz decomposition \eqref{Lefshetz}.

We now consider the four-fold $CY_4=K3 \times \widetilde{K3}$. The middle form cohomology splits as 
\begin{equation}
H^4(CY_4) = \left[ H^0(K3) \otimes H^4(\tK)\right]  \, \oplus \, \left[ H^2(K3) \otimes H^2(\tK)\right] \, \oplus \,  \left[ H^4(K3) \otimes H^0(\tK)\right] \,,
\end{equation}
where the first and the last factors have dimension one. Fluxes on these 4-cycles wrap either one or the other $K3$ and can in principle stabilize their relative volume. Since these are not allowed in F-theory, we restrict ourselves to
\begin{equation}\label{eq:k3flux}
G_4 \in H^2(K3, \bbZ) \otimes H^2(\tK, \bbZ) \,,
\end{equation}
which are fluxes with two legs on each of the two $K3$.

This choice can stabilize all the geometric moduli of each $K3$ except for their volumes. There are $2\times (58-1)$ real geometric moduli, or $57$ complex moduli.\footnote{As explained in Section \ref{sec:modstab}, the primitivity condition on $G_4$ is non-trivial on $K3\times K3$ and it gives conditions that allow to fix the K\"ahler moduli. Furthermore, in this section we do not require $G_4$ to have a good F-theory limit and, as such, the flux \eqref{eq:k3flux} can stabilize all moduli except for the volume of either $K3$.} The self-duality condition \eqref{sd4} then implies
\begin{equation}
G_4 \in H^2_+(K3) \otimes H^2_+(\tK) \; \oplus \; H^2_-(K3) \otimes H^2_-(\tK) \,.
\end{equation}

To understand how the four-form flux \eqref{eq:k3flux} stabilizes the moduli of $K3\times\tK$, ref.~\cite{Braun:2008pz} introduces the homomorphisms $g \colon H^2(\tK) \rightarrow H^2(K3)$ and its adjoint (with respect to the inner product \eqref{K3prod}) $\tilde g \colon H^2(K3) \rightarrow H^2(\tK)$, defined by
\begin{equation}\label{eq:gandgtilde}
g (\tilde v) = \int_\tK G_4 \wedge \tilde v  \,,\qquad \tilde g (v) = \int_{K3} G_4 \wedge v \,.
\end{equation}
If one expands $G_4$ with respect to the integer bases of $H^2(K3)$ and $H^2(\tK)$,
\begin{equation}
G_4 = N^{i \tilde \jmath} \alpha_i \wedge \tilde \alpha_{\tilde \jmath} \,,
\end{equation}
the corresponding matrix representations of $g$ and $\tilde g$ are
\begin{equation}
g^i{}_{\tilde \jmath} = N^{i \tilde k} d_{\tilde k \tilde \jmath} \,,\qquad \tilde g^{\tilde \imath}{}_j = (N^T)^{\tilde \imath k} d_{kj} \ .
\end{equation}
One can now rewrite the potential \eqref{eq:potential} as
\begin{equation}\label{eq:k3k3potential}
V = - \cV^{-3} \left(\sum_{\hat \imath} \Bigl\| \tilde P_-\bigl[\tilde g (\omega_{\hat \imath}) \bigr]\Bigr\|^2 + \sum_i \Bigl\| P_- \bigl[g (\tilde \omega_{\hat \imath}) \bigr]\Bigr\|^2 \right) \,,
\end{equation}
where $P_-$ and $\tilde P_-$ denote the projectors onto $H^2_-(K3)$ and $H^2_-(\tK)$ respectively.
To obtain a Minkowski minimum both terms must vanish independently which happens if and only if
\begin{equation}\label{eq:minkcond}
g (H^2_+(\tK)) \subset H^2_+(K3) \,, \qquad \tilde g (H^2_+(K3)) \subset H^2_+(\tK) \,.
\end{equation}
This is equivalent to the fact that $g$ and $\tilde g$ map positive-norm vectors into positive-norm vectors (not necessarily bijectively).
This condition is equivalent to requiring the self-adjoint map $\tilde g \circ g\colon H^2(\tK) \rightarrow H^2(\tK)$ to be diagonalizable with real, non-negative eigenvalues.
Since $\tilde g$ is the adjoint of $g$, \eqref{eq:minkcond} is equivalent to
\begin{equation}\label{eq:minkcondb}
g ( H^2_-(\tK)) \subset H^2_-(K3) \,, \qquad \tilde g ( H^2_-(K3)) \subset H^2_-(\tK) \,.
\end{equation}
Therefore, \eqref{eq:minkcond} implies that $\tilde g g$ is block-diagonal.
Since each block is self-adjoint with respect to definite metrics, $\tilde g g$ is diagonalizable with non-negative eigenvalues.
As explained in Appendix \ref{App:moreK3K3}, if there are two equal eigenvalues whose eigenvectors have positive norm and negative norm, then there is a flat direction.
Moreover, if $H^2_-(K3)$ or $H^2_-(\tK)$ contains an integer vector of norm $-2$ (a root of the K3 lattice), the respective K3 has a shrinking cycle corresponding to an orbifold singularity.
We will therefore exclude these possibilities from our search.

Let us summarize the conditions on smooth compactifications giving Minkowski minima (they are further explained in Appendix~\ref{App:moreK3K3}):
\begin{itemize}
\item There exists a Minkowski minimum iff the matrices $(g\tilde g)^i{}_j = N^{i \tilde k} d_{\tilde k \tilde l} N^{m\tilde l} d_{m j}$ and $(\tilde g g)^{\tilde \imath}{}_{\tilde \jmath} = N^{k \tilde \imath} d_{kl} N^{l\tilde m} d_{\tilde m \tilde \jmath}$ are diagonalizable with non-negative eigenvalues.%
\footnote{This does a priori not exclude vacua with zero eigenvalues.
However, when we have a degenerate zero eigenvalue and the corresponding eigenspace contains vectors of both positive and negative norm, there are situations where no vacuum exists even though all eigenvalues are non-negative.
This exception is excluded by the condition on moduli stabilization and is therefore not relevant for our discussion.
For further details see Appendix~\ref{App:moreK3K3}.
}
\item
The eigenvectors of $g\tilde g$ with positive norm span $H^2_+(K3)$ and those with negative norm span $H^2_-(K3)$.
Therefore, they determine a point in the moduli space of $K3$.
The equivalent statement holds for the eigenvectors of  $\tilde g g$ and the moduli space of $\tK$.
\item Denote the eigenvalues of positive or negative norm eigenvectors by $\{a_1,a_2,a_3\}$ and $\{b_1, \dots, b_{19}\}$, respectively.
If there is no $a_i$ and $b_j$ such that $a_i = b_j$ all moduli are stabilized. Otherwise the potential has flat directions.
\item {If there are no vectors such that}
\begin{equation}
v \in H^2_-(K3) \cap H^2(K3, \mathbb{Z}) \,,\qquad (v,v) = -2 \,,
\end{equation}
then $K3$ does not have an orbifold singularity.\footnote{{Singular compactifications on $K3 \times K3$, where the moduli are stabilized at locations corresponding to orbifold singularities were found in \cite{Dasgupta:1999ss,Aspinwall:2005ad,Braun:2008pz,Braun:2014ola,Kimura:2016gxw}}} Equivalently for $\tK$.
\item The contribution to the tadpole cancellation condition is given by
\begin{equation}
\frac12 \int G_4 \wedge G_4 = \tfrac12 N^{i \tilde k}  d_{\tilde k \tilde l} N^{n\tilde l} d_{n i} = \tfrac12 \mathrm{tr} \, g \tilde g \,.
\end{equation}
\end{itemize}
Moreover, the minimum is supersymmetric if $a_1 = a_2$ and $a_3 = 0$.

\end{subsection}

\begin{subsection}{Differential evolution search}\label{sec:desearch}

Here we just give a brief outline of the search and present the results.
The details will be given in the companion paper \cite{algorithm}.

To make a best effort attempt at finding a minimum tadpole there are a multitude of global search algorithms at our disposal, and we have opted for one known as \emph{differential evolution}. This is a population-based algorithm where the population is evolved towards a minimum of a \emph{cost-} or \emph{fitness-}function using evolutionary operations; mutation, crossover, and selection.
The problem we aim to solve is finding a matrix with the lowest tadpole possible satisfying the conditions on a smooth flux-compactification as described above.
The operations of differential evolution act on a population of $N$ real vectors that define candidate solutions. Here, the population consists of $22 \times 22 = 484$ dimensional vectors, $x \in \mathbb{R}^{484}$, that when rounded to integers represent the $484$ entries of the flux matrix. Differential evolution then applies the following operations on individual vectors taken from a population of such input vectors.

The mutation operation is performed on an input vector, $x$, to form a new vector, $x'$, according to
\begin{equation}
  x' = x + \sum_{\alpha=1}^{n} F^\alpha (y_\alpha - z_\alpha)\,,
\end{equation}
where $\{y_\alpha,z_\alpha\}_{\alpha = 1,\ldots,n}$, are pairs of other members in the population, $n$ being some choice done during implementation (commonly $n = 1$ or $2$). The numbers $F^{\alpha}$ are known as \emph{differential weights}, and act as a sort of step-size.

The crossover operation is when we select only some of the entries of $x'$ to replace those of $x$, forming a vector $x''$
\begin{equation}
  {x''}_i = \left\{ \begin{array}{ll} {x'}_i & {\texttt{if rand()}} < C_r \\ x_i \end{array} \right.
\end{equation}
The number $C_r$ is called a \emph{crossover probability} and affects the direction of the evolution.

The final step would be to compare the original $x$ and the newly derived $x''$ in some way to determine (possibly in a comparison with other such pairs) which vector will stay in the population. This is done in what it called a selection operation. For selection to take place each vector must be assigned a number, called fitness, that determines how well it solves the goals of Section \ref{sec:desearch}. 
We can design a fitness-function by computing a weighted sum of penalties, where each penalty is some measure of how far we are to the goal, and the weights determine the priority of each goal. The penalties we have designed are
\begin{itemize}
  \item Eigenvalue penalties: Negative or complex eigenvalues of $g\tilde g$ receive a penalty depending on how far away they are from being positive.\footnote{For negative eigenvalues it is simply the absolute value of the eigenvalue, for complex numbers it is split in two parts: absolute value of the real part if negative, and the absolute value of the imaginary part.}
  \item Diagonalization: There is a constant penalty if $g\tilde g$ or $\tilde g g$ is found to be non-diagonalizable, which is computed by comparing the algebraic and geometric multiplicities.
\item Degeneracy of eigenvalues: If two or more eigenvalues of $g\tilde g$ are the same, their corresponding eigenvectors $e_i$ form a bilinear form $b_{ij} = (e_i,e_j)$, and the penalty is equal to the number of negative or positive eigenvalues of this $b_{ij}$, whichever is the smallest.
  \item Norm-2 vectors: Our fitness function has several methods for attempting to find the number of integer norm 2 vectors (roots) in $H^2_-(K3)$ in different sets. This is a NP-hard problem and hence a fast way of determining their presence is not available. We use two different approaches that utilize a lattice-reduction algorithm inspired by the Lenstra Lenstra Lovasz (LLL) algorithm \cite{Lenstra82factoringpolynomials}. If any of these methods finds a norm 2 vector, a constant number is given as the penalty.
  \item Tadpole: { The induced charge $\frac12 \mathrm{tr} \, g \tilde g$ is given as the penalty.}\footnote{{In the absence of D3 branes (M2 branes in the M-theory picture) tadpole cancellation is satisfied for $\frac12 \mathrm{tr} \, g \tilde g = 24$.
However, here we simply search for flux configurations with minimal induced charge and thus also allow $\frac12 \mathrm{tr} \, g \tilde g < 24$ (when tadpole cancelations can be satisfied by including additional branes).}}
\end{itemize}
The weights are then chosen such that the algorithm will solve all penalties (by finding solutions where they are all zero) except the tadpole, which the differential evolution algorithm attempts to minimize.

We apply a differential evolution algorithm using \texttt{BlackBoxOptim.jl} \cite{Feldt2018},\footnote{There are several differential evolution algorithms provided by \texttt{BlackBoxOptim.jl}, we use the default ``adaptive DE/rand/1/bin with radius limited sampling''.} which is a package for the Julia programming language \cite{Julia-2017}. The fitness-function we have designed will be made available as part of the \texttt{bbsearch.jl} repository of the companion paper \cite{algorithm}.

The search does not {\em guarantee} to find the global minima of the fitness-function, but differential evolution is a search algorithm designed for this purpose. This being said, as we will discuss in \cite{algorithm}, we believe that the apparent convergence of our searches is a strong indication of the existence of a minimum bound for the tadpole, within solutions satisfying the desired properties.

The final result of the differential evolution search is then supplemented by a local search that turns entries on and off in the matrices found to see if the result can be improved. We will describe this method in more detail in \cite{algorithm}.

\subsection{Result of the search}

 The minimum $Q^{\text{flux}}_{\text{D3}}$ that we find using the differential evolution and local search is
\begin{equation}
  Q^{\text{flux}}_{\text{D3}} = 25\,.
\end{equation}
This minimum was reached by several independent searches. We employed two different strategies:
\begin{itemize}
  \item Large-population ($> 1000$) searches over long time ($\sim$ weeks)
  \item Small-population ($\le 1000$) searches over short time ($\sim$ days)
\end{itemize}
A larger population means a larger part of the search space is explored, hence can increase the chances of finding the global minima, but it takes longer to get there, while smaller populations easier fall into local minima quickly. We explored both strategies to hedge our bets.
After a local search for each population, most searches resulted in matrices with tadpole 25 which satisfied all our goals. In total we have generated $\gtrsim 10^4$ matrices of tadpole 25, in searches that considered $\gtrsim 10^8$ matrices, and not a single matrix with tadpole 24 or lower was found. One of our verified matrices with tadpole 25 can be found in Appendix \ref{app:matrix}.

The minimum found is larger than the allowed tadpole
\beq
\frac{\chi_{K3\times K3}}{24}=24
\eeq
and indicates that we cannot stabilize all 57 complex ``stabilizable" moduli of $K3\times K3$  (all moduli except the overall volumes of each $K3$) away from singularities, within the tadpole bound. The minimum value of the flux-tadpole constant, $\alpha$, our search produces is\footnote{As explained in the previous section, in this example the number 57 counts all moduli which can be stabilized by fluxes.}
\beq
\alpha= \frac{25}{57}\approx 0.439
\eeq
comfortably satisfying the conjectured inequality \eqref{alpha}.
\end{subsection}

\section{Moduli stabilization in Type IIB compactifications with D7-branes: an example}
\label{sec:D7}

In this section we show that the problem of stabilizing complex-structure moduli in F-theory compactifications is related in Type IIB String Theory to the problem of stabilizing D7-brane moduli. Indeed, the $h^{3,1}(Z)$ complex-structure moduli in F-theory compactifications on a four-fold $Z$ reduce in the Type IIB limit to $h^{2,1}_-(X)$ complex-structure moduli of the 3-fold $X$, plus $h_-^{(2,0)}(S)$ deformations of  the cycle $S$ wrapped by D7-branes. Both these types of moduli need to be stabilized by adding (bulk or brane) fluxes, at the price of increasing the tadpole. We review in this section the example of moduli stabilization worked out by Collinucci, Denef and  Esole in \cite{Collinucci:2008pf}, and link it to our Conjecture~\hyperlink{conjecture3b}{3b}.

This example is based on one of the simplest elliptically fibered four-folds, $Z$, given by an elliptic fibration over ${\mathbb C}{\mathbb P}^3$. This is described by the usual Weierstrass equation
\beq
Z \colon \ y^2=x^3+f(u) \, x \, z^4 + g(u) \, z^6 \,,
\eeq
where $f(u)$, $g(u)$ are homogeneous polynomials of degree 16 and 24 respectively of the base coordinates $(u_1,u_2,u_3,u_4)\in {\mathbb C}{\mathbb P}^3$, and $(x,y,z)$ are the coordinates on the fiber. The Euler and Hodge numbers of the four-fold are
\beq
\chi(Z)=23328 \ , \quad h^{1,1}(Z)=2\ , \quad h^{2,1}(Z)=2\ , \quad h^{2,2}(Z)=15564\ , \quad h^{3,1}(Z)=3878\ .
\eeq
In the Sen limit \cite{Sen:1996vd, Sen:1997gv}, the polynomials $f$ and $g$ take the form
\beq\begin{aligned}
f&=-3h^2+\epsilon \eta \ , \\
g&=-2h^3 + \epsilon h \eta - \epsilon^2 \chi / 12 \,,
\end{aligned}\eeq
where $h$, $\eta$ and $\chi$ are homogeneous polynomials of degree 8, 16 and 24. The Type IIB limit is $\epsilon \to 0$, where the coupling is weak everywhere except near $h=0$, which corresponds to the locations of the O7-planes. The D7-branes are located at
\beq
{\rm D7} \colon \ \eta^2(u)=h(u) \, \chi(u) \ .
\eeq
The D7 deformation moduli are  the number of inequivalent deformations of this degree 32 equation in the variables $(u_1,u_2,u_3)$, given by
\beq \label{ND7mod}
N_{\rm D7~mod}=\begin{pmatrix} 16+3 \\ 3 \end{pmatrix} +\begin{pmatrix} 24+3 \\ 3 \end{pmatrix} - \begin{pmatrix} 8+3 \\ 3 \end{pmatrix} -1
=3728
\eeq
where the first subtraction is from the equivalence $(\eta,\chi)\simeq (\eta+ h \psi , \chi + 2 \eta \psi+h \psi^2)$ and the -1 corresponds to an overall rescaling of the coefficients. Note that the 3878 complex-structure moduli of the four-fold correspond mostly to D7-moduli in the Type IIB limit.

These moduli have to be fixed by D7 worldvolume fluxes. These fluxes are given by primitive (1,1) forms, Poincar\'e dual to holomorphic curves $\gamma$ on the divisor $S$ wrapped by the D7-branes\footnote{Note that $S$ is the de-singularized version of the divisor $S$ wrapped by the D7-branes, where the pinch-point singularities at the points where the D7 branes intersect the O7 planes have been blown up.}. For a curve of degree $d$, the number of constraints resulting from requiring it to be on the D7-branes is
\beq \label{Nconstr}
N_{\rm constr}(\gamma)= 32 d+1
\eeq
Since we want all moduli fixed, one requires
\beq \label{dge}
N_{\rm constr} \ge N_{\rm D7~mod} \Rightarrow d\ge 117  \ .
\eeq
On the other hand, the induced D3-charge of the fluxes is given by
 \bea \label{QD3F}
 Q_{\rm D3}(F)&=&-{\color{black} \frac12} \int_{S}  F^2= \frac{\chi(\gamma)}{\color{black} 2} - {\color{black}\frac12} \int_{S} \gamma \cdot c_1(S) \nonumber \\
 &=&\frac{\chi(\gamma)}{\color{black} 2}+{\color{black} 14} d \ge {\color{black} 14} d
\eea
where in the second equality Collinucci, Denef and  Esole used the adjunction formula \eqref{adj}. The computation of $c_1(S)$ is in Eq.~(3.34) of \cite{Collinucci:2008pf}.  The last inequality holds for $\chi(\gamma) >0$ (and if $\chi(\gamma)$ is slightly negative the second term would still dominate\footnote{The situation for large negative $\chi(\gamma)$ is unclear as there are also curve moduli to take into account.}).
Putting everything together we have
\beq
Q_{\rm D3}(F)\ge {\color{black} 14 }d \ge {\color{black} 1638} > \frac{\chi}{24} = {\color{black} 972}
\eeq
and thus D7 moduli cannot be fixed within the tadpole.

It is instructive to see how the D3 tadpole induced by the fluxes needed to stabilize moduli behaves as a function of the number of moduli. For that,  we write the number of D7 moduli, Eq. \eqref{ND7mod}, and the induced D3-charge of the fluxes needed to stabilize them, Eq. \eqref{QD3F}, as a function of the number of D7-branes $m$ ($m=16$ in \eqref{ND7mod} and \eqref{QD3F})
\bea
N_{\rm D7~mod}&=&\small { \begin{pmatrix} m+3 \\ 3 \end{pmatrix} +\begin{pmatrix} 2m-5 \\ 3 \end{pmatrix} - \begin{pmatrix} m-5 \\ 3 \end{pmatrix}}-1=\frac43 m^3-8m^2+\frac{59}{3} m \nonumber \\
 Q_{\rm D3}(F)&=&\frac{\chi(\gamma)}{\color{black} 2}+({\color{black} m-2})d \ge {\color{black} \frac23 m^3 -\frac{16}{3} m^2 + \frac{107}{6} m-\frac{59}{3}}
 \eea
 where in the last inequality we have used the bound on $d$ in \eqref{dge} for generic $m$:
  \beq
 d\ge \frac23 m^2 -4m +\frac{59}{6},
 \eeq
 ignoring the +1 term in \eqref{Nconstr} which is independent of $m$ and subleading. We see that for large $m$, the D3-charge sourced by the fluxes needed to stabilize all D7 moduli grows precisely as {\color{black} half} the number of moduli, which means that, for this example, the flux-tadpole constant in Conjecture~\hyperlink{conjecture3b}{3b} is
 \beq
 \alpha \approx 1/2
 \label{alphaCol}
 \eeq

Let us point out certain key features of this example, some of which are generic and likely to apply to other compactifications with D7 branes wrapping cycles with a large Euler number.  The important factors that enter in determining the flux-tadpole constant, $\alpha$, are the scaling  of the number of constraints and of the induced D3 charge with respect to the number of D7 moduli, $m$. For large $m$ the former scales like $\tfrac16 (2m)^3$ just because the location of the D7-branes is defined by an equation of degree $2m$ of three variables, which are the coordinates of $\mathbb {CP}^3$.
The number of constraints \eqref{Nconstr}  scales like $2md$. This gives a universal bound $d\ge \tfrac23 m^2$ at leading order.
 Finally, the scaling of the induced D3-charge depends on $c_1(S)$, which goes like $\color{black} -m$. These behaviors appear to be model independent, which indicates that  the D3-charge sourced by the worldvolume fluxes needed to stabilize D7 moduli is generically equal to about half the number of moduli. We leave the formalization of this argument for future work.

 \section{Tadpoles bounds and de Sitter constructions}

 \label{KS}

The most controlled way to obtain de Sitter solutions in String Theory involves three steps: stabilizing the complex-structure moduli, stabilizing the K\"ahler moduli, and uplifting the cosmological constant of the resulting AdS compactification by the addition of supersymmetry-breaking anti-D3 branes \cite{Kachru:2003aw}. In order to preserve the stabilization of the moduli achieved at the second stage, the positive contribution to the cosmological constant brought about by the antibranes should be very small, and therefore they can only be placed in a region of the compactification manifold where the warp factor is large.

The prototypical example of such a region is the warped deformed conifold solution constructed by Klebanov and Strassler \cite{Klebanov:2000hb}. This solution has $M$ units of RR three-form flux wrapping the three-cycle corresponding to the deformation of the conifold, as well as an NSNS three-form field strength wrapping a non-compact three-cycle that contains the two sphere that collapses at the tip of the conifold and the radial direction. When the KS throat is embedded in a flux compactification this cycle is also compact, and is wrapped by $K$ units of NSNS three-form flux. This flux determines the length of the KS throat and the ratio of $K$ and $M$ controls the hierarchy between the bottom of the throat and the rest of the compactification,
\beq
\frac{\Lambda_{IR}}{\Lambda_{UV}} = \exp\left(-{ 2 \pi K \over 3 g_s M}\right) \,.
 \eeq
This hierarchy has to be of order $10^{-10} \approx e^{-23}$  if we want the anti-D3 branes used for uplifting to preserve the stability of the K\"ahler moduli \cite{Kachru:2003aw}. This implies that
\beq
2 \pi K > 23 \times 3 g_s M \,.
\label{Kbound}
\eeq

In a previous paper, three of the authors and Duda\c s have shown that adding even a single antibrane to the KS throat will destabilize the complex-structure modulus corresponding to the deformation of the conifold,\footnote{In \cite{Randall:2019ent} this instability was interpreted as the destabilization of the radion in a five-dimensional RS model \cite{Randall:1999ee} with mismatched IR brane tension.} unless the RR flux on this cycle,
 \beq
 { M > 6.8~\sqrt{N_{\overline{D3}} \over g_s}}.
\label{Mbound}
\eeq
%
%
In \cite{Bena:2019sxm}, two of the authors and Buchel have compared this to the bound on the existence of Klebanov-Strassler black holes \cite{Buchel:2018bzp} and also found a bound with the same functional form in a regime of parameters where backreaction is important.\footnote{One could also try to derive a bound on the conifold flux by extrapolating the probe calculation of  \cite{Kachru:2002gs} to argue that anti-D3 branes polarizing into a single NS5 brane are unstable for $M>12/N_{\overline{D3}}$ and hence do not give rise to metastable minima. This bound was recently interpreted  as a bound on $M$, by taking $N_{\overline{D3}}=1$~\cite{Gao:2020xqh}. We do not believe this interpretation is justified for two reasons: i) single branes do not undergo brane polarization \cite{Myers:1999ps} and ii) according to the calculation in \cite{Kachru:2002gs} the probe anti-D3-branes can also polarize into multiple ($n_5$) NS5-branes when $M>\tfrac{12}{ N_{\overline{D3}}\, n_5}$. Hence this calculation cannot be used to put a lower bound on $M$. }

 Equations \eqref{Kbound} and \eqref{Mbound} imply that the contribution of the fluxes in the KS throat to the D3 brane tadpole is
\beq
Q_{\rm D3}^{\rm throat} \equiv
K M > { 23 \times 3 \times 6.8^2\over 2 \pi} \approx 500 \,. \label{stability}
\eeq

One of the questions that has motivated our investigation is whether one can cancel such a large tadpole contribution in a flux compactification with stabilized moduli. As we have seen in this paper, obtaining a negative contribution to the D3 tadpole is not hard. In the language of F-theory this can be done by considering four-folds with a large Euler number, and in the language of Type IIB compactifications this can be done using D7 branes  that wrap four-cycles with a large Euler number.

However, negative tadpole always comes with the price of extra moduli, which need to be stabilized using fluxes. The main result of our calculations has been that these fluxes induce in turn a large positive contribution to the D3 tadpole, which is much larger than the absolute value of the negative tadpole we wanted to obtain in the first place.

Hence, if one wants to cancel a tadpole of order 500, the options are very limited, and possibly non-existent. From a Type IIB perspective the negative-charge ingredients we have are O3 planes, O7 planes and D7 branes. Since the number of O3 involutions of a six-dimensional manifold is limited, the largest amount of tadpole one can hope to cancel using them is $64/{\color{black}4= 16}$.\footnote{We thank Andre Lukas and David Morrison for discussions on this topic.} This is largely insufficient for our needs.

Therefore, the only hope to obtain a sufficiently large negative tadpole to cancel $Q_{\rm D3}^{\rm throat}$ comes from O7 planes and D7 branes. Although these come hand-in-hand in order to cancel D7-brane tadpole, let us for the moment consider them separately. The O7 planes do not have associated moduli, and hence one may na\"ively hope that using O7 planes would be the best strategy. However, there are two problems:  First, the maximum negative tadpole one can hope to obtain from O7 planes is bounded above by \eqref{O7bound2} for all the  Calabi-Yau three-folds from the Kreuzer-Skarke list \cite{Kreuzer:2000xy}, and this bound is equal to one third of the total number of K\"ahler and complex-structure moduli. Since one can only stabilize a few K\"ahler moduli\footnote{For an interesting recent exploration of the interaction of K\"ahler moduli stabilization with the existence of a long warped throat see \cite{Gao:2020xqh}.}, and since stabilizing complex-structure moduli has to be done with fluxes which source D3 charge, one can use Conjecture~\hyperlink{conjecture3a}{3a} to argue that these fluxes will generically give a larger positive contribution to the D3 tadpole than the negative contribution of the O7 planes. 

Furthermore, as mentioned above, O7 planes must be accompanied by D7 branes, whose moduli have to be stabilized with D7 worldvolume fluxes that increase the D3 tadpole. The D7 branes can either sit on top of the O7 planes, or wrap a more complicated divisor. When the D7 branes and the O7 planes are coincident, $\chi(D7) = 8 \chi(O7)$ and one can use again the Lefshetz fixed-point theorem to show that the absolute value of their negative contribution to the D3 tadpole is bounded above. 

However, as we have seen in Section~\ref{sec:D7}, when the D7 branes are moved away from the O7 planes, then $\chi(D7)$ can become much much larger than $ 8 \chi(O7)$, and hence can give a very large negative D3-tadpole contribution.  However, from \eqref{eq:D7moduli} we can see that a large $\chi(D7)$ always comes with a large number of D7 moduli, which have to be stabilized using D7 worldvolume fluxes. Conjecture~\hyperlink{conjecture3b}{3b} in turn implies that these fluxes give a large positive induced D3 charge. 
The ratio between the positive induced charge and the negative tadpole contribution in the large-$\chi(D7)$  limit is
\beq
{Q_{\rm D3}^{\rm D7 \,moduli\, stabilization} \over \left| Q^{-}_{\displaystyle \chi_{\rm D7}} \right|} \approx { \alpha n_{\rm moduli} \over {n_{\rm moduli}/4} } = 4 \alpha \,,
\eeq
If one uses the value of $\alpha$ corresponding to the particular example discussed in Section \ref{sec:D7}, we can easily see that trying to increase the absolute value of the negative D3 tadpole by increasing the Euler characteristic of the four-cycle wrapped by D7 branes, one has to pay back four times more induced D3 charge if one wants to stabilize the resulting D7 moduli, so this is always a losing game.

As we can see, all the negative tadpole one can hope to obtain will be over-run by the positive induced charge coming from the brane fluxes needed to fix the D7 moduli, and hence canceling 500 units of positive charge does not appear possible. This in turn indicates that uplifting with anti-D3 branes placed in long Klebanov-Strassler-like throats cannot be done while stabilizing all the other moduli of the compactification.

 \section{Discussion}

\label{sec:disc}

 There are several important questions raised by our investigation. The first, and most important one, is whether one can prove the conjectures we made. All the evidence we have gathered so far supports them. For example, the fact that we could generate 35000 tadpole-25 solutions that stabilize all the stabilizable moduli of $K3 \times K3$, but not even a single tadpole-24 solution, points clearly to the existence of a lower bound on the charge of the fluxes needed to stabilize  moduli.

This being said, a clear calculation demonstrating this bound would be an amazing step. Setting up this calculation is complicated, both in F-theory (Conjecture~\hyperlink{conjecture1}{1}) and in Type IIB String Theory (Conjecture \hyperlink{conjecture3a}{3a}), since the stabilization of the moduli involves (besides topological data such as intersection numbers) the Hodge duals of the cycles, which are in general a complicated function of the moduli. In our upcoming companion paper \cite{algorithm} we will present further arguments that illustrate the issues with this stabilization.

On the other hand, arriving at a proof of Conjecture~\hyperlink{conjecture3b}{3b} in the large-Euler-characteristic limit appears to be more feasible. Indeed, as we have seen in Section \ref{sec:D7}, this  involves more or less straightforward D7-brane physics and arguments that appear rather general. Once one establishes this conjecture, it would be interesting to try to use its F-theory translation to see if it provides a root to proving Conjecture~\hyperlink{conjecture1}{1} as well.

The leading-order behavior we have found clearly excludes both F-theory compactifications with a large number of stable complex-structure moduli, as well as compactifications where one can cancel the 500 units of tadpole necessary for anti-D3 uplift. The second important question is whether the growth of the tadpole with the number of stabilized moduli we have conjectured for large numbers of moduli can somehow be contorted by subleading corrections.

There are several strategies one could use to try to squeeze these bounds. One may hope to squeeze a few dozen negative units of tadpole from subleading corrections, or from increasing the number of K\"ahler moduli in \eqref{chiCY4} as much as possible. However, as we have discussed in Section \ref{KS}, to stabilize these K\"ahler moduli one needs D7 branes with gaugino condensation, whose number is bound by D7 tadpole-cancelation conditions. One can also try to use all the 64 possible O3 planes. However, we can see that these numbers are rather small.

As far as anti-D3 brane uplift is concerned, one might also try to decrease a bit the $10^{10}$ hierarchy between scale at the bottom of the KS throat the scale of the CY compactification.\footnote{Decreasing this hierarchy too much will destabilize the K\"ahler moduli, and also result in an unphysically large gravitino mass.~\cite{Dudas:2019pls}. See also \cite{Blumenhagen:2019qcg} for a discussion of the anti-D3 brane instability in the context of K\"ahler moduli stabilization and the swampland distance conjecture.} One may also hope that the bound on $M$ in equation \eqref{Mbound} (obtained in \cite{Bena:2018fqc} using the corrections computed in \cite{Douglas:2008jx}) could be slightly modified by corrections. One may also try to uplift using an anti-D3 brane on top of an O3 plane \cite{Antoniadis:1999xk,Kallosh:2015nia}, which has half the mass of a normal anti-D3 brane and may lower the bound on $M$ and on the tadpole contribution of the warped throats that allow for such configurations \cite{Garcia-Etxebarria:2015lif}. None of these steps is guaranteed to work, but their existence leaves a tiny glimmer of hope that antibrane uplift might still give a few reasonable metastable de Sitter vacua in some very limited corner of the string compactification landscape. We believe that all these directions are worth exploring if one is to understand where de Sitter vacua could be hiding in the Landscape.

This being said, both Conjecture~\hyperlink{conjecture1}{1}, which rules out the possibility of F-theory compactifications with a large number of stabilized moduli, and the arguments above, which show that antibrane uplift is very non-generic and may only be possible in an extremely small sliver of the string landscape, indicate that all estimations of the number of String Theory  compactifications with stabilized moduli, including numbers like $10^{500}$ \cite{Denef:2004ze}  or $10^{272000}$ \cite{Taylor:2015xtz}, have to be taken with a (boulder-sized) grain of salt.

\vskip 1cm

\noindent {\bf Acknowledgements:}
 We would like to thank Andreas Braun, Duiliu Diaconescu, Emilian Duda\c s, Thomas Grimm, Jim Halverson, Arthur Hebecker, Andre Lukas, David Morrison, Jakob Moritz, Savdeep Sethi and Cumrun Vafa for valuable discussions.
 This work was supported in part by the ANR grant Black-dS-String ANR-16-CE31-0004-01, by the ERC Grants 772408 ``Stringlandscape'' and 787320 ``QBH Structure'', by the John Templeton Foundation grant 61149 and by the US National Science Foundation Grant PHY-1748958.  M.G. would like to thank the KITP for hospitality while part of this work was carried out. The work of J.B.~was supported by MIUR-PRIN contract 2015MP2CX4002 ``Non-perturbative aspects of gauge theories and strings''.
The work of S.L.~was supported by the US National Science Foundation grant PHY-1915071.
Parts of this work were performed at the Aspen Center for Physics, which is supported by National Science Foundation grant PHY-1607611.


\appendix


\section{F-theory and IIB moduli}
\label{App:moduli}

 In this table taken from \cite{Denef:2008wq} we list the F-theory moduli and give their Type IIB origin.

\hspace*{-.7cm}\begin{tabular}[h]{llll}
 F-theory/M-theory & number of real moduli & IIB with O3 and O7 & number of real moduli \\
 \hline
 K\"ahler & $h^{1,1}(Z)-1$ & K\"ahler & $h^{1,1}_+(X)$ \\ \\
 && Complex structure & $2 \, h^{2,1}_-(X)$ \\
 Complex structure & $2 \, h^{3,1}(Z)$ & D7 deformations & $2 \, \hat{h}^{2,0}_-(S)$ \\
 && Dilaton-axion & 2 \\ \\
 $C_6$ axions & $h^{1,1}(Z)-1$ & $C_4$ axions & $h^{1,1}_+(X)$ \\ \\
 $C_3$ axions & $2 \, h^{2,1}(Z)$ & $B_2$, $C_2$ axions & $h^{1,1}_-(X) + h^{1,1}_-(X)$ \\ \\
 M2 positions & $6 \, N_{\rm D3}$ & D3 positions & $6 \, N_{\rm D3}$
\end{tabular}
\vskip5mm
The subscripts $\pm$ denote the Hodge numbers counting the even and respectively the odd parts of the relevant cohomology under the orientifold involution.

\section{Details of moduli stabilization on $K3 \times K3$}
\label{App:moreK3K3}

In this Appendix we give further details on the vacuum conditions used in Section~\ref{K3xK3}, mainly following \cite{Braun:2008pz}.

The homomorphisms $g \colon H^2(\tK) \rightarrow H^2(K3)$ and  $\tilde g \colon H^2(K3) \rightarrow H^2(\tK)$, defined in \eqref{eq:gandgtilde} define two endomorphisms $g \tilde g \colon H^2(K3) \rightarrow H^2(K3)$ and $\tilde g  g \colon H^2(\tK) \rightarrow H^2(\tK)$.
Assume that these endomorphisms are diagonalizable with non-negative eigenvalues.
Since $\tilde g g$ is self-adjoint, there exists a complete basis of orthogonal eigenvectors.
Notice that this basis can be chosen in such a way that all eigenvectors have non-vanishing norm.%
\footnote{To see this, assume there exists an eigenvector, $v_0$, such that $\|v_0\| = 0$. Then there must exist another eigenvector, $v'$, such that $(v_0, v') \neq 0$ because otherwise there would not be a complete basis of eigenvectors or the inner product would be degenerate.
For selfadjoint matrices eigenvectors with different eigenvalues are orthogonal.
Hence, it follows that $v'$ must have the same eigenvalue of $v_0$ and we can construct two new eigenvectors of positive and negative norm as linear combinations of $v_0$ and $v'$.}
This defines a decomposition of $H^2(\tK)$ into $H^2_+(\tK)$ and $H^2_-(\tK)$ such that $H^2_{\pm}(\tK)$ are spanned by the positive- and negative-norm eigenvectors of $\tilde g g$ respectively.
Let $\tilde v \in H^2(\tK)$ be an eigenvector of $\tilde g g$ with non-vanishing eigenvalue $\lambda > 0$ and $v = g \tilde v \in H^2(K3)$.
Then $v$ is an eigenvector of $g \tilde g$ with the same eigenvalue $\lambda$.
This shows that $g \tilde g$ and $\tilde g g$ have the same spectrum.
Moreover, $\|v\|^2 = (g \tilde v, g \tilde v) = (\tilde g g \tilde v, \tilde v) = \lambda \|\tilde v\|^2$, so $g$ preserves the sign of the norm.

However, if $\tilde v$ has eigenvalue $\lambda = 0$ then the situation is slightly more subtle.
Here it follows from the same argument as above that  $\| g \tilde v \|^2 = 0$.
This allows for two different possibilities:

The first is that, $v = g \tilde v \neq 0$.%
\footnote{Ref.~\cite{Braun:2008pz} wrongly argues that this situation cannot occur.
They claim that since $g$ and $\tilde g$ are adjoint, there is the orthogonal decomposition $H^2(K3) = \mathrm{Im}\, g \oplus \mathrm{Ker} \,\tilde g$.
Then, for $\tilde v$ such that $\tilde g g\tilde v = 0$ one has $g \tilde v \in\mathrm{Im}\,{g} \cap \mathrm{Ker}\,{\tilde g} = \{0\}$ and hence $g \tilde v = 0$.
However, this argument is not correct on spaces of indefinite signature.
Here is a simple counter example:
Take $d = \left(\begin{smallmatrix}1 & 0 \\ 0 & -1 \end{smallmatrix}\right)$, $N = \left(\begin{smallmatrix}1 & 1 \\ 1 & 1 \end{smallmatrix}\right)$ and therefore $g  = N d = \tilde g = N^T d = \left(\begin{smallmatrix}1 & -1 \\ 1 & -1 \end{smallmatrix}\right)$.
Consequently, $\tilde g g =  \left(\begin{smallmatrix}0 & 0 \\ 0 & 0 \end{smallmatrix}\right)$ has a two-fold degenerate zero eigenvalue.
Take now one of the corresponding eigenvectors $\tilde v = (1,\, 0)^T$ such that $v = g \tilde v = (1,\, 1)^T$.
Clearly, $\tilde g v = 0$ and thus $v \in \mathrm{Im}\,{g} \cap \mathrm{Ker}\,{\tilde g}$ but $v \neq 0$.
Notice, that this is only possible because $\| v \|^2 = 0$ which requires the kernel of $g \tilde g$ to be of indefinite signature.}
Then $g$ maps a vector of non-vanishing norm onto a null vector and either equation \eqref{eq:minkcond} or or equation \eqref{eq:minkcondb} is violated.
Therefore, the potential \eqref{eq:k3k3potential} has no minimum.
Moreover, $gv =0$ and therefore the null vector $v$ is an eigenvector of $\tilde g g$ with eigenvalue zero.
This implies that this problematic situation can only occur if the eigenvalue zero is degenerate and if the corresponding eigenspace has indefinite signature.

The second possibility is that all vectors $\tilde v$ in the kernel of $\tilde g g$ satisfy $g \tilde v = 0$ (which implies they are also in the kernel of $g$). However, these vectors  cannot be in the image of $\tilde g$ because otherwise they were null (we argued above that we can always choose a basis of non-null eigenvectors).
Therefore, for each such $\tilde v$ there also exists a $v \in H^2(K3)$ such that $\tilde g v = 0$.
This shows that the eigenvectors of $g \tilde g$ are given by the direct sum of the kernel of $\tilde g$ and the image of the eigenvectors of $\tilde g g$ under $g$.
Therefore, we can define $H^2_{\pm}(K3)$ to be spanned by the positive- and negative-norm eigenvectors of $g \tilde g$ and equation \eqref{eq:minkcond} holds.

The decomposition of $H^2(\tK)$ into eigenspaces of positive and negative norm is not always unique.
Assume that there are two eigenvectors, $\tilde v_+$ and $\tilde v_-$, of positive and negative norm respectively, such that they have the same eigenvalue.
Then we can define
\begin{equation}
\tilde v'_+ = \tilde v_+ + \epsilon \tilde v_- \,,
\end{equation}
which is again an eigenvector of positive norm if $\epsilon$ is sufficiently small.
Therefore, in this case the decomposition is not unique.
in physics terms the potential has a flat direction and not all moduli are stabilized.
Notice that this problem does not arise if two of the positive norm eigenvectors or two of the negative norm eigenvectors have the same eigenvalues.

The previous conditions are not enough to guarantee the existence of a supersymmetric minimum.
For this we have to additionally impose \eqref{susyconditions}.
Primitivity of the flux implies
\begin{equation}
g \tilde \omega_3 = \tilde g \omega_3 = 0 \,,
\end{equation}
and this implies that there must be a positive norm vector in the kernel of $g$ and $\tilde g$ (one implies the other).
On the other hand, the absence of a (4,0) or (0,4) component implies
\begin{equation}
0 = (\omega, g \tilde \omega) = (\omega_1, g \tilde\omega_1) - (\omega_2, g \tilde \omega_2) \,,
\end{equation}
and hence the remaining two positive-norm eigenvectors have equal eigenvalues.

In general, the eigenvectors of $g\tilde g$ and $\tilde g g$ are non-integer and thus do not intersect with the lattices $H^2(K3, \mathbb{Z})$ or $H^2(\tK, \mathbb{Z})$.
However, let us assume that there is a vector
\begin{equation}\label{eq:intvector}
v \in H^2_-(K3) \cap H^2(K3, \mathbb{Z}) \,,
\end{equation}
which is equivalent to assuming that there is a linear combination of negative-norm eigenvectors of $g \tilde g$ which is integer. Let us furthermore assume that
\begin{equation}
(v,v) = -2 \,,
\end{equation}
so $v$ is a root of the lattice $H^2(K3, \mathbb{Z})$.
As $H^2_+(K3)$ and $H^2_-(K3)$ are orthogonal, $v$ is orthogonal to $H^2_+(K3)$.
However, the existence of a root orthogonal to the three-plane $H^2_+(K3)$ implies that we are at a special point on the $K3$ moduli space with an orbifold singularity (see for example \cite{Aspinwall:1996mn}).
The same statement holds for roots in $H^2_-(\tK)$ and orbifold singularities of $\tK$.

Let us finally note that there is nothing special about eigenvalues $\lambda = 0$. However, if $\lambda \in \mathbb{Z}$ (which includes $\lambda = 0$) the corresponding eigenvector, $v$, can be chosen to satisfy \eqref{eq:intvector} because it is in the kernel of the integer matrix $g \tilde g - \lambda \mathbb{1}$.
This does however not yet guarantee that $(v,v) = -2$.
Moreover, vectors satisfying \eqref{eq:intvector} need not have zero
integer eigenvalues. In general, whenever the characteristic polynomial of  $g \tilde g$ factorizes over the integers and when the roots of one of the irreducible factors correspond to eigenvalues with only negative-norm eigenvectors, there are integer vectors satisfying \eqref{eq:intvector}.
Further details on this can be found in \cite{algorithm}.

\section{Example matrix}
\label{app:matrix}

This is one example of a matrix what has $Q^{\text{flux}}_{\text{D3}} = 25$,
\begingroup\makeatletter\def\f@size{8}\check@mathfonts
$$ \left[N^{i\tilde \jmath}\right] = \left[
\begin{array}{cccccccccccccccccccccc}
 0 & 0 & 0 & 0 & -1 & 0 & 0 & 0 & 0 & 0 & 0 & 0 & 0 & 0 & 0 & 0 & 0 & 0 & 0 & 0 & 0 & 0 \\
0 & 0 & 0 & 0 & 0 & -1 & 0 & 0 & 0 & 0 & 0 & 0 & 0 & 0 & 0 & 0 & 0 & 0 & 0 & 0 & 0 & 0 \\
0 & 0 & 0 & 0 & 0 & 0 & 0 & 0 & 0 & 0 & 0 & -1 & 0 & 0 & -1 & 1 & 0 & 0 & 0 & 0 & 0 & 0 \\
-1 & 0 & 0 & 0 & 0 & 0 & 0 & 0 & 0 & 0 & 0 & 0 & 0 & 0 & 0 & 0 & 0 & 0 & 0 & 0 & 0 & 0 \\
0 & 0 & 0 & 1 & 0 & 0 & 0 & 0 & 0 & 0 & 0 & 0 & 0 & 0 & 1 & 0 & -1 & 0 & 0 & 0 & 0 & 0 \\
0 & 0 & 0 & 0 & 0 & 0 & 0 & 0 & 0 & 0 & 0 & 0 & 0 & 0 & -1 & 0 & 0 & 0 & 0 & 0 & 0 & 0 \\
0 & 0 & 0 & 0 & 0 & 0 & 0 & 0 & 0 & 1 & 1 & 1 & 0 & 0 & 0 & 0 & 0 & 0 & 0 & 0 & 0 & 0 \\
-1 & 0 & 0 & 0 & 0 & 0 & -1 & -1 & -1 & 0 & 0 & 0 & 0 & 0 & 0 & 0 & 0 & 0 & 0 & 0 & 0 & 0 \\
0 & 0 & -1 & 1 & 0 & 0 & 0 & 0 & 0 & 0 & 0 & 0 & 0 & 0 & 0 & 0 & 0 & 0 & 0 & 0 & 0 & 0 \\
1 & 0 & 0 & 0 & 0 & 0 & 0 & 0 & 0 & 0 & 0 & 0 & 0 & 0 & 0 & 0 & 0 & 0 & 0 & 0 & 0 & 0 \\
0 & 0 & 0 & 0 & 1 & -1 & 0 & 0 & 0 & 1 & 1 & 0 & 0 & 1 & 0 & 0 & 0 & 0 & 0 & 0 & 0 & 0 \\
0 & 0 & 0 & 0 & 0 & 0 & 0 & 0 & 0 & 0 & 0 & -1 & 0 & 0 & 0 & 0 & 0 & 0 & 0 & 0 & 0 & 0 \\
0 & 0 & 0 & 0 & 0 & 0 & 0 & 0 & 0 & 0 & 0 & 0 & 0 & 0 & 0 & 0 & 0 & 0 & 0 & 0 & 0 & 0 \\
0 & 0 & 0 & 0 & 1 & -1 & 0 & 0 & 0 & 0 & 0 & 0 & 0 & 0 & 0 & 0 & 0 & 0 & 0 & 0 & 0 & 0 \\
0 & 0 & 0 & 0 & 0 & 0 & 0 & 0 & 0 & 0 & 0 & 0 & 0 & 0 & 0 & 0 & 0 & 1 & 1 & 0 & 0 & 1 \\
1 & 0 & 0 & 0 & 0 & 0 & 0 & 0 & 0 & 0 & 0 & 0 & 0 & 0 & 0 & 0 & 0 & 0 & 0 & 0 & 0 & 1 \\
1 & 0 & 0 & 0 & 0 & 0 & 0 & 0 & 0 & 0 & 0 & 0 & 0 & 0 & 0 & 0 & 0 & -1 & -1 & -1 & 0 & 0 \\
0 & 0 & 0 & 0 & 0 & 0 & 0 & 0 & 0 & 0 & 0 & 0 & 0 & 0 & 0 & 0 & 0 & 0 & 0 & 0 & 0 & 0 \\
0 & 1 & 1 & 0 & 0 & 0 & 1 & 1 & 0 & 0 & 0 & 0 & 0 & 0 & 0 & 0 & 0 & 0 & 1 & 0 & 0 & 1 \\
-1 & 0 & 0 & 0 & 0 & 0 & 0 & 1 & 0 & 0 & 0 & 0 & 0 & 0 & 0 & -1 & 0 & 0 & 1 & 0 & 0 & 0 \\
0 & 0 & 0 & 0 & 0 & 0 & 0 & 0 & 0 & 0 & 0 & 0 & 0 & 0 & 0 & -1 & 0 & 0 & 0 & 0 & 0 & 0 \\
0 & 0 & 1 & 0 & 0 & 0 & 0 & 0 & -1 & -1 & 0 & 0 & 0 & 0 & 0 & 0 & 0 & 0 & 0 & 0 & 0 & 0
\end{array}
\right] \,. $$
\endgroup
For this choice of fluxes both matrices $g \tilde g = N d N^{T} d$ and $\tilde g g = N^T d N d$ are diagonalizable and have the eigenvalues
\begin{equation}\begin{gathered}
1, 0.293, 0.164 \,, \\
7.93,7.52,5.91,5.32,4.71,4.34,3.12,2.19,1.95,1.56,1.49,1.05,0.774,0.676,0,0,0,0,0 \,,
\end{gathered}\end{equation}
where those in the first line have eigenvectors of positive norm and those in the second line have eigenvectors of negative norm.
The sum of the eigenvalues is $2 \times 25$.
There are no negative eigenvalues, therefore this flux matrix would give rise to a potential with a (non-supersymmetric) Minkowski minimum.
Notice, that the only degenerate eigenvalue is ``0''.
However, the corresponding eigenspaces (of $g\tilde g$ and $\tilde g g$) have negative definite signature and thus all the moduli are stabilized.
Moreover, since ``0'' is an integer eigenvalue, its eigenspaces can be spanned by integer vectors (lattice elements), but it can be checked that they do not contain any roots (integer vectors of norm -2).
Therefore, the compactification is smooth.
The only condition which is not satisfied is tadpole cancellation.



\bibliographystyle{JHEP}
\bibliography{refs-tadpole}

\providecommand{\href}[2]{#2}\begingroup\raggedright\begin{thebibliography}{10}

\bibitem{Dasgupta:1999ss}
K.~Dasgupta, G.~Rajesh, and S.~Sethi, {\it {M theory, orientifolds and G -
  flux}},  {\em JHEP} {\bf 08} (1999) 023,
  [\href{http://arxiv.org/abs/hep-th/9908088}{{\tt hep-th/9908088}}].

\bibitem{Collinucci:2008pf}
A.~Collinucci, F.~Denef, and M.~Esole, {\it {D-brane Deconstructions in IIB
  Orientifolds}},  {\em JHEP} {\bf 02} (2009) 005,
  [\href{http://arxiv.org/abs/0805.1573}{{\tt arXiv:0805.1573}}].

\bibitem{Braun:2020jrx}
A.~P. Braun and R.~Valandro, {\it {$G_4$ Flux, Algebraic Cycles and Complex
  Structure Moduli Stabilization}},
  \href{http://arxiv.org/abs/2009.11873}{{\tt arXiv:2009.11873}}.

\bibitem{Grana:2001xn}
M.~{Gra\~na} and J.~Polchinski, {\it {Gauge / gravity duals with holomorphic
  dilaton}},  {\em Phys. Rev.} {\bf D65} (2002) 126005,
  [\href{http://arxiv.org/abs/hep-th/0106014}{{\tt hep-th/0106014}}].

\bibitem{Giddings:2001yu}
S.~B. Giddings, S.~Kachru, and J.~Polchinski, {\it {Hierarchies from fluxes in
  string compactifications}},  {\em Phys. Rev.} {\bf D66} (2002) 106006,
  [\href{http://arxiv.org/abs/hep-th/0105097}{{\tt hep-th/0105097}}].

\bibitem{Gukov:1999ya}
S.~Gukov, C.~Vafa, and E.~Witten, {\it {CFT's from Calabi-Yau four folds}},
  {\em Nucl. Phys. B} {\bf 584} (2000) 69--108,
  [\href{http://arxiv.org/abs/hep-th/9906070}{{\tt hep-th/9906070}}]. [Erratum:
  Nucl.Phys.B 608, 477--478 (2001)].

\bibitem{Taylor:2015xtz}
W.~Taylor and Y.-N. Wang, {\it {The F-theory geometry with most flux vacua}},
  {\em JHEP} {\bf 12} (2015) 164, [\href{http://arxiv.org/abs/1511.03209}{{\tt
  arXiv:1511.03209}}].

\bibitem{Kachru:2003aw}
S.~Kachru, R.~Kallosh, A.~D. Linde, and S.~P. Trivedi, {\it {De Sitter vacua in
  string theory}},  {\em Phys. Rev.} {\bf D68} (2003) 046005,
  [\href{http://arxiv.org/abs/hep-th/0301240}{{\tt hep-th/0301240}}].

\bibitem{Denef:2004ze}
F.~Denef and M.~R. Douglas, {\it {Distributions of flux vacua}},  {\em JHEP}
  {\bf 05} (2004) 072, [\href{http://arxiv.org/abs/hep-th/0404116}{{\tt
  hep-th/0404116}}].

\bibitem{Demirtas:2019sip}
M.~Demirtas, M.~Kim, L.~Mcallister, and J.~Moritz, {\it {Vacua with Small Flux
  Superpotential}},  {\em Phys. Rev. Lett.} {\bf 124} (2020), no.~21 211603,
  [\href{http://arxiv.org/abs/1912.10047}{{\tt arXiv:1912.10047}}].

\bibitem{Demirtas:2020ffz}
M.~Demirtas, M.~Kim, L.~Mcallister, and J.~Moritz, {\it {Conifold Vacua with
  Small Flux Superpotential}},  \href{http://arxiv.org/abs/2009.03312}{{\tt
  arXiv:2009.03312}}.

\bibitem{Blumenhagen:2020ire}
R.~\'Alvarez-Garc\'\i{}a, R.~Blumenhagen, M.~Brinkmann, and L.~Schlechter, {\it
  {Small Flux Superpotentials for Type IIB Flux Vacua Close to a Conifold}},
  \href{http://arxiv.org/abs/2009.03325}{{\tt arXiv:2009.03325}}.

\bibitem{Lust:2019zwm}
D.~{L\"ust}, E.~Palti, and C.~Vafa, {\it {AdS and the Swampland}},
  \href{http://arxiv.org/abs/1906.05225}{{\tt arXiv:1906.05225}}.

\bibitem{Alday:2019qrf}
L.~F. Alday and E.~Perlmutter, {\it {Growing Extra Dimensions in AdS/CFT}},
  {\em JHEP} {\bf 08} (2019) 084, [\href{http://arxiv.org/abs/1906.01477}{{\tt
  arXiv:1906.01477}}].

\bibitem{Klebanov:2000hb}
I.~R. Klebanov and M.~J. Strassler, {\it {Supergravity and a confining gauge
  theory: Duality cascades and chi SB resolution of naked singularities}},
  {\em JHEP} {\bf 08} (2000) 052,
  [\href{http://arxiv.org/abs/hep-th/0007191}{{\tt hep-th/0007191}}].

\bibitem{Bena:2018fqc}
I.~Bena, E.~Dudas, M.~{Gra\~na}, and S.~{L\"ust}, {\it {Uplifting Runaways}},
  {\em Fortsch. Phys.} {\bf 67} (2019), no.~1-2 1800100,
  [\href{http://arxiv.org/abs/1809.06861}{{\tt arXiv:1809.06861}}].

\bibitem{Denef:2005mm}
F.~Denef, M.~R. Douglas, B.~Florea, A.~Grassi, and S.~Kachru, {\it {Fixing all
  moduli in a simple f-theory compactification}},  {\em Adv. Theor. Math.
  Phys.} {\bf 9} (2005), no.~6 861--929,
  [\href{http://arxiv.org/abs/hep-th/0503124}{{\tt hep-th/0503124}}].

\bibitem{Honma:2017uzn}
Y.~Honma and H.~Otsuka, {\it {On the Flux Vacua in F-theory
  Compactifications}},  {\em Phys. Lett. B} {\bf 774} (2017) 225--228,
  [\href{http://arxiv.org/abs/1706.09417}{{\tt arXiv:1706.09417}}].

\bibitem{Blaback:2013ht}
J.~Bl\r{a}b\"ack, U.~Danielsson, and G.~Dibitetto, {\it {Fully stable dS vacua
  from generalised fluxes}},  {\em JHEP} {\bf 08} (2013) 054,
  [\href{http://arxiv.org/abs/1301.7073}{{\tt arXiv:1301.7073}}].

\bibitem{Damian:2013dq}
C.~Damian, L.~R. Diaz-Barron, O.~Loaiza-Brito, and M.~Sabido, {\it {Slow-Roll
  Inflation in Non-geometric Flux Compactification}},  {\em JHEP} {\bf 06}
  (2013) 109, [\href{http://arxiv.org/abs/1302.0529}{{\tt arXiv:1302.0529}}].

\bibitem{Damian:2013dwa}
C.~Damian and O.~Loaiza-Brito, {\it {More stable de Sitter vacua from S-dual
  nongeometric fluxes}},  {\em Phys. Rev. D} {\bf 88} (2013), no.~4 046008,
  [\href{http://arxiv.org/abs/1304.0792}{{\tt arXiv:1304.0792}}].

\bibitem{Blaback:2013fca}
J.~Bl\r{a}b\"ack, U.~Danielsson, and G.~Dibitetto, {\it {Accelerated Universes
  from type IIA Compactifications}},  {\em JCAP} {\bf 03} (2014) 003,
  [\href{http://arxiv.org/abs/1310.8300}{{\tt arXiv:1310.8300}}].

\bibitem{Blaback:2013qza}
J.~Bl\r{a}b\"ack, D.~Roest, and I.~Zavala, {\it {De Sitter Vacua from
  Nonperturbative Flux Compactifications}},  {\em Phys. Rev. D} {\bf 90}
  (2014), no.~2 024065, [\href{http://arxiv.org/abs/1312.5328}{{\tt
  arXiv:1312.5328}}].

\bibitem{Abel:2014xta}
S.~Abel and J.~Rizos, {\it {Genetic Algorithms and the Search for Viable String
  Vacua}},  {\em JHEP} {\bf 08} (2014) 010,
  [\href{http://arxiv.org/abs/1404.7359}{{\tt arXiv:1404.7359}}].

\bibitem{Ruehle:2017mzq}
F.~Ruehle, {\it {Evolving neural networks with genetic algorithms to study the
  String Landscape}},  {\em JHEP} {\bf 08} (2017) 038,
  [\href{http://arxiv.org/abs/1706.07024}{{\tt arXiv:1706.07024}}].

\bibitem{Cole:2019enn}
A.~Cole, A.~Schachner, and G.~Shiu, {\it {Searching the Landscape of Flux Vacua
  with Genetic Algorithms}},  {\em JHEP} {\bf 11} (2019) 045,
  [\href{http://arxiv.org/abs/1907.10072}{{\tt arXiv:1907.10072}}].

\bibitem{AbdusSalam:2020ywo}
S.~AbdusSalam, M.~Cicoli, F.~Quevedo, P.~Shukla, and S.~Abel, {\it {A
  systematic approach to K\"ahler moduli stabilisation}},  {\em JHEP} {\bf 08}
  (2020), no.~08 047, [\href{http://arxiv.org/abs/2005.11329}{{\tt
  arXiv:2005.11329}}].

\bibitem{CaboBizet:2020cse}
N.~Cabo~Bizet, C.~Damian, O.~Loaiza-Brito, D.~K.~M. Pe\~na, and J.~Monta\~nez
  Barrera, {\it {Testing Swampland Conjectures with Machine Learning}},  {\em
  Eur. Phys. J. C} {\bf 80} (2020), no.~8 766,
  [\href{http://arxiv.org/abs/2006.07290}{{\tt arXiv:2006.07290}}].

\bibitem{Betzler:2019kon}
P.~Betzler and E.~Plauschinn, {\it {Type IIB flux vacua and tadpole
  cancellation}},  {\em Fortsch. Phys.} {\bf 67} (2019), no.~11 1900065,
  [\href{http://arxiv.org/abs/1905.08823}{{\tt arXiv:1905.08823}}].

\bibitem{Ruehle:2020jrk}
F.~Ruehle, {\it {Data science applications to string theory}},  {\em Phys.
  Rept.} {\bf 839} (2020) 1--117.

\bibitem{algorithm}
I.~Bena, J.~Bl{\aa}b{\"a}ck, M.~{Gra\~na}, and S.~{L\"ust}, {\it {Genetic
  Algorithm for the Tadpole Problem}},  \href{http://arxiv.org/abs/to
  appear}{{\tt to appear}}.

\bibitem{Bena:2005va}
I.~Bena and N.~P. Warner, {\it {Bubbling supertubes and foaming black holes}},
  {\em Phys. Rev.} {\bf D74} (2006) 066001,
  [\href{http://arxiv.org/abs/hep-th/0505166}{{\tt hep-th/0505166}}].

\bibitem{Berglund:2005vb}
P.~Berglund, E.~G. Gimon, and T.~S. Levi, {\it {Supergravity microstates for
  BPS black holes and black rings}},  {\em JHEP} {\bf 06} (2006) 007,
  [\href{http://arxiv.org/abs/hep-th/0505167}{{\tt hep-th/0505167}}].

\bibitem{Bena:2006is}
I.~Bena, C.-W. Wang, and N.~P. Warner, {\it {The Foaming three-charge black
  hole}},  {\em Phys. Rev.} {\bf D75} (2007) 124026,
  [\href{http://arxiv.org/abs/hep-th/0604110}{{\tt hep-th/0604110}}].

\bibitem{Giryavets:2003vd}
A.~Giryavets, S.~Kachru, P.~K. Tripathy, and S.~P. Trivedi, {\it {Flux
  compactifications on Calabi-Yau threefolds}},  {\em JHEP} {\bf 04} (2004)
  003, [\href{http://arxiv.org/abs/hep-th/0312104}{{\tt hep-th/0312104}}].

\bibitem{Denef:2004dm}
F.~Denef, M.~R. Douglas, and B.~Florea, {\it {Building a better racetrack}},
  {\em JHEP} {\bf 06} (2004) 034,
  [\href{http://arxiv.org/abs/hep-th/0404257}{{\tt hep-th/0404257}}].

\bibitem{Klemm:1996ts}
A.~Klemm, B.~Lian, S.~Roan, and S.-T. Yau, {\it {Calabi-Yau fourfolds for M
  theory and F theory compactifications}},  {\em Nucl. Phys. B} {\bf 518}
  (1998) 515--574, [\href{http://arxiv.org/abs/hep-th/9701023}{{\tt
  hep-th/9701023}}].

\bibitem{Kreuzer:2000xy}
M.~Kreuzer and H.~Skarke, {\it {Complete classification of reflexive polyhedra
  in four-dimensions}},  {\em Adv. Theor. Math. Phys.} {\bf 4} (2002)
  1209--1230, [\href{http://arxiv.org/abs/hep-th/0002240}{{\tt
  hep-th/0002240}}].

\bibitem{Maldacena:1997de}
J.~M. Maldacena, A.~Strominger, and E.~Witten, {\it {Black hole entropy in M
  theory}},  {\em JHEP} {\bf 12} (1997) 002,
  [\href{http://arxiv.org/abs/hep-th/9711053}{{\tt hep-th/9711053}}].

\bibitem{Denef:2008wq}
F.~Denef, {\it {Les Houches Lectures on Constructing String Vacua}},  {\em Les
  Houches} {\bf 87} (2008) 483--610,
  [\href{http://arxiv.org/abs/0803.1194}{{\tt arXiv:0803.1194}}].

\bibitem{Haack:2001jz}
M.~Haack and J.~Louis, {\it {M theory compactified on Calabi-Yau fourfolds with
  background flux}},  {\em Phys. Lett. B} {\bf 507} (2001) 296--304,
  [\href{http://arxiv.org/abs/hep-th/0103068}{{\tt hep-th/0103068}}].

\bibitem{Sen:1996vd}
A.~Sen, {\it {F theory and orientifolds}},  {\em Nucl. Phys.} {\bf B475} (1996)
  562--578, [\href{http://arxiv.org/abs/hep-th/9605150}{{\tt hep-th/9605150}}].

\bibitem{Sen:1997gv}
A.~Sen, {\it {Orientifold limit of F theory vacua}},  {\em Phys. Rev. D} {\bf
  55} (1997) 7345--7349, [\href{http://arxiv.org/abs/hep-th/9702165}{{\tt
  hep-th/9702165}}].

\bibitem{Aspinwall:2005ad}
P.~S. Aspinwall and R.~Kallosh, {\it {Fixing all moduli for M-theory on
  K3xK3}},  {\em JHEP} {\bf 10} (2005) 001,
  [\href{http://arxiv.org/abs/hep-th/0506014}{{\tt hep-th/0506014}}].

\bibitem{Braun:2008pz}
A.~P. Braun, A.~Hebecker, C.~Ludeling, and R.~Valandro, {\it {Fixing D7 Brane
  Positions by F-Theory Fluxes}},  {\em Nucl. Phys.} {\bf B815} (2009)
  256--287, [\href{http://arxiv.org/abs/0811.2416}{{\tt arXiv:0811.2416}}].

\bibitem{Braun:2010ff}
A.~P. Braun, {\em {F-Theory and the Landscape of Intersecting D7-Branes}}.
\newblock PhD thesis, Heidelberg U., 2010.
\newblock \href{http://arxiv.org/abs/1003.4867}{{\tt arXiv:1003.4867}}.

\bibitem{Braun:2014ola}
A.~P. Braun, Y.~Kimura, and T.~Watari, {\it {The Noether-Lefschetz problem and
  gauge-group-resolved landscapes: F-theory on K3 $\times$ K3 as a test case}},
   {\em JHEP} {\bf 04} (2014) 050, [\href{http://arxiv.org/abs/1401.5908}{{\tt
  arXiv:1401.5908}}].

\bibitem{Kimura:2016gxw}
Y.~Kimura, {\it {Gauge symmetries and matter fields in $\mathrm{F}$-theory
  models without section \textemdash{} compactifications on double cover and
  Fermat quartic $\mathrm{K}3$ constructions times $\mathrm{K}3$}},  {\em Adv.
  Theor. Math. Phys.} {\bf 21} (2017) 2087--2114,
  [\href{http://arxiv.org/abs/1603.03212}{{\tt arXiv:1603.03212}}].

\bibitem{Lenstra82factoringpolynomials}
A.~K. Lenstra, H.~W. Lenstra, and L.~Lovasz, {\it Factoring polynomials with
  rational coefficients},  {\em MATH. ANN} {\bf 261} (1982) 515--534.

\bibitem{Feldt2018}
R.~Feldt, ``Blackboxoptim.jl.''
  \url{https://github.com/robertfeldt/BlackBoxOptim.jl}, 2018.

\bibitem{Julia-2017}
J.~Bezanson, A.~Edelman, S.~Karpinski, and V.~B. Shah, {\it Julia: A fresh
  approach to numerical computing},  {\em SIAM {R}eview} {\bf 59} (2017), no.~1
  65--98.

\bibitem{Randall:2019ent}
L.~Randall, {\it {The Boundaries of KKLT}},  {\em Fortsch. Phys.} {\bf 68}
  (2020), no.~3-4 1900105, [\href{http://arxiv.org/abs/1912.06693}{{\tt
  arXiv:1912.06693}}].

\bibitem{Randall:1999ee}
L.~Randall and R.~Sundrum, {\it {A Large mass hierarchy from a small extra
  dimension}},  {\em Phys. Rev. Lett.} {\bf 83} (1999) 3370--3373,
  [\href{http://arxiv.org/abs/hep-ph/9905221}{{\tt hep-ph/9905221}}].

\bibitem{Bena:2019sxm}
I.~Bena, A.~Buchel, and S.~{L\"ust}, {\it {Throat destabilization (for profit
  and for fun)}},  \href{http://arxiv.org/abs/1910.08094}{{\tt
  arXiv:1910.08094}}.

\bibitem{Buchel:2018bzp}
A.~Buchel, {\it {Klebanov-Strassler black hole}},  {\em JHEP} {\bf 01} (2019)
  207, [\href{http://arxiv.org/abs/1809.08484}{{\tt arXiv:1809.08484}}].

\bibitem{Kachru:2002gs}
S.~Kachru, J.~Pearson, and H.~L. Verlinde, {\it {Brane / flux annihilation and
  the string dual of a nonsupersymmetric field theory}},  {\em JHEP} {\bf 06}
  (2002) 021, [\href{http://arxiv.org/abs/hep-th/0112197}{{\tt
  hep-th/0112197}}].

\bibitem{Gao:2020xqh}
X.~Gao, A.~Hebecker, and D.~Junghans, {\it {Control issues of KKLT}},
  \href{http://arxiv.org/abs/2009.03914}{{\tt arXiv:2009.03914}}.

\bibitem{Myers:1999ps}
R.~C. Myers, {\it {Dielectric branes}},  {\em JHEP} {\bf 12} (1999) 022,
  [\href{http://arxiv.org/abs/hep-th/9910053}{{\tt hep-th/9910053}}].

\bibitem{Dudas:2019pls}
E.~Dudas and S.~{L\"ust}, {\it {An update on moduli stabilization with
  antibrane uplift}},  \href{http://arxiv.org/abs/1912.09948}{{\tt
  arXiv:1912.09948}}.

\bibitem{Blumenhagen:2019qcg}
R.~Blumenhagen, D.~Kl{\"a}wer, and L.~Schlechter, {\it {Swampland Variations on
  a Theme by KKLT}},  {\em JHEP} {\bf 05} (2019) 152,
  [\href{http://arxiv.org/abs/1902.07724}{{\tt arXiv:1902.07724}}].

\bibitem{Douglas:2008jx}
M.~R. Douglas and G.~Torroba, {\it {Kinetic terms in warped
  compactifications}},  {\em JHEP} {\bf 05} (2009) 013,
  [\href{http://arxiv.org/abs/0805.3700}{{\tt arXiv:0805.3700}}].

\bibitem{Antoniadis:1999xk}
I.~Antoniadis, E.~Dudas, and A.~Sagnotti, {\it {Brane supersymmetry breaking}},
   {\em Phys. Lett. B} {\bf 464} (1999) 38--45,
  [\href{http://arxiv.org/abs/hep-th/9908023}{{\tt hep-th/9908023}}].

\bibitem{Kallosh:2015nia}
R.~Kallosh, F.~Quevedo, and A.~M. Uranga, {\it {String Theory Realizations of
  the Nilpotent Goldstino}},  {\em JHEP} {\bf 12} (2015) 039,
  [\href{http://arxiv.org/abs/1507.07556}{{\tt arXiv:1507.07556}}].

\bibitem{Garcia-Etxebarria:2015lif}
I.~Garcia-Etxebarria, F.~Quevedo, and R.~Valandro, {\it {Global String
  Embeddings for the Nilpotent Goldstino}},  {\em JHEP} {\bf 02} (2016) 148,
  [\href{http://arxiv.org/abs/1512.06926}{{\tt arXiv:1512.06926}}].

\bibitem{Aspinwall:1996mn}
P.~S. Aspinwall, {\it {K3 surfaces and string duality}},  in {\em {Theoretical
  Advanced Study Institute in Elementary Particle Physics (TASI 96): Fields,
  Strings, and Duality}}, pp.~421--540, 11, 1996.
\newblock \href{http://arxiv.org/abs/hep-th/9611137}{{\tt hep-th/9611137}}.

\end{thebibliography}\endgroup

\end{document}